\documentclass[11pt,a4paper]{article}
\pdfoutput=1
\usepackage{exscale,times}
\usepackage{fancyhdr}
\usepackage[cp1250]{inputenc}
\usepackage[T1]{fontenc}
\usepackage{amsmath}
\usepackage{fancyhdr}
\usepackage{indentfirst}
\usepackage{times}
\usepackage{graphicx}
\usepackage{multicol}
\usepackage{here}
\usepackage{amsmath}
\usepackage{amssymb}
\usepackage{amsthm}
\usepackage{amsfonts}
\usepackage{stmaryrd}
\usepackage{eufrak}
\usepackage{mathrsfs}
\usepackage{color}
\usepackage{multicol}
\usepackage{mathrsfs}
\usepackage{wasysym}

\newcommand{\bta}{\mbox{\boldmath $\tau$}}

\newcommand{\bsigma}{\mbox{\boldmath $\sigma$}}

\newcommand{\bveps}{\mbox{\boldmath $\varepsilon$}}

\numberwithin{equation}{section}

\begin{document}
\pagestyle{fancy}

\chead{Wojciech Sumelka - August 2013}
\lhead{}
\rhead{}
\renewcommand{\headrulewidth}{0.4pt}

\begin{center}
%
%   Please insert the title here...
%
{\fontsize{14}{20}\bf{Non-local fractional model of rate independent plasticity }}
\end{center}

\begin{center}
\textbf{Wojciech Sumelka}\\
\bigskip
{Poznan University of Technology, Institute of Structural Engineering, }\\
Piotrowo 5 street, 60-969 Poznan, Poland\\ 
wojciech.sumelka@put.poznan.pl\\
\bigskip
\end{center}
%
%   Please insert the keywords here...
%


\vspace{1cm}

\begin{center}
\textbf{ABSTRACT}\\[1mm]
\end{center}
In the paper the generalisation of classical rate independent plasticity using fractional calculus is presented. This new formulation is non-local due to properties of applied fractional differential operator during definition of kinematics. In the description small fractional strains assumption is hold together with additive decomposition of total fractional strains into elastic and plastic parts. Classical local rate independent plasticity is recovered as a special case.

%\tableofcontents

\section{Introduction}

Experimental observations show that matter, independently on state, has in general characteristic length scale. Thus, a measurable quantity (e.g. strain) in a particular point of interest include somehow information from its surrounding. Considering modelling, proper mathematical description of such experimental facts should include both: (i) the length scale, and (ii) a law which governs the way in which the information from the surrounding  affect the material point. Such models we call non-local.

In this paper we consider solid bodies which reveal the ability to handle permanent deformation (e.g. steel, concrete, rubber). We propose new non-local model for their behaviour description. Namely, we utilise generalised version of classical continuum mechanics, called \textit{fractional continuum mechanics} \cite{Sumelka2013-work2013,Sumelka2013-CMM2013}. This model is stated in terms of fractional calculus - an individual branch of pure mathematics in which differential operators of arbitrary order (even complex) are considered. This new formulation manifests significant flexibility, in sense of possible experimental results approximation, together with the positive feature (in contrast to classical non-local models \cite{Sumelka2012-CM,Eftis2003-IJP}) that only two additional material parameters are included.

It should be pointed out, that the applications of fractional calculus in physics (since its invention in 1695 \cite{Leibniz62}) has found recently a great attention among scientist  e.g. in Fluid Flow, Rheology, Dynamical Processes in Self-Similar and Porous Structures, Diffusive Transport Akin to Diffusion, Electrical Networks, Probability and Statistics, Control Theory of Dynamical Systems, Viscoelasticity, Electrochemistry of Corrosion, Chemical Physics, Optics, and others cf. \cite{Samko93,Podlubny1999-AP,Leszczynski2005,Kilbas06,Tarasov2008, Mainardi2010,Leszczynski2011}. However the application to the plasticity has not been recorded to the authors knowledge, thus to some extend, this paper opens new perspectives for plasticity theory.

The proposed generalisation of classical continuum mechanics using fractional calculus exists in the literature (cf. \cite{Vazaquez2004,Lazopoulos2006,Paola2009,Atanackovic2009,Carpinteri2011,Drapaca2012}). However the description presented in this paper has at least the following important original features: (i) the proposed new formulation has clear physical interpretation and is developed by analogy to general framework of classical continuum mechanics; (ii) we deal with finite deformations (in contrast to \cite{Atanackovic2009,Carpinteri2011} where small deformations are considered only); (iii) contrary to previous works e.g. \cite{Atanackovic2009,Carpinteri2011,Drapaca2012} the generalised fractional measures of the deformation e.g. fractional deformation gradients or fractional strains have the same physical dimensions as classical one (thus their classical interpretation remain unchanged); (iv) characteristic length scale of the particular material is defined explicitly (an in classical non-local models); (v) objectivity requirements are proved; (vi) and finally, the discussed concept bases on the fractional material and spatial line elements in contrast to \cite{Drapaca2012} where the whole formulation bases on fractional motion (what can be important because in more general formulations displacement field may not exist cf. \cite{MarsdenHughes83} Box 3.1 pp 57). Thus, as a conclusion, the discussed way of introducing non-locality to the description is new, and extends the classical techniques  i.e. explicit \cite{Borst96-IJNME,Fleck97-AAM,Aifantis99-IJF,Voyiadjis2005,Polizzotto2011} (e.g. via classical strain gradients) or implicit \cite{Perzyna1998,Lodygowski1997a,Dorn2002c,Sumelka2011-AAM2011,Sumelka2012-CM,Voyiadjis2013-IJIE,Sumelka2013-JEM-ASCE} (i.e. via relaxation time in Perzyna's type viscoplasticity).

The paper is structured as follows. In Section 2 non-local fractional continua is defined. Section 3 deals with non-local fractional model of rate independent plasticity. In Section 4 illustrative examples together with precise numerical flow chart are presented.

\section{Non-local fractional kinematics}

\subsection{Riesz-Caputo fractional derivative}

Let us consider the Riesz-Caputo (RC) fractional derivative, the one used for further definition of fractional continua. We have for a function $f(t)$  ($t\in(a,b)\subseteq \mathbb{R}$ and $0<\alpha < 1$ - when $\alpha$ is an integer, the usual definition of a derivative is used) \cite{Agrawal2007,Frederico2010}

\begin{equation}\label{eq:RC}
~_{~~a}^{RC}D^{\alpha}_b f(t)=\frac{1}{2}\frac{\Gamma(2-\alpha)}{\Gamma(2)}\left(~_{a}^{C}D^{\alpha}_t f(t)+(-1)^n~_{t}^{C}D^{\alpha}_b  f(t)\right),	
\end{equation}

where $\alpha>0$ denotes the real order of the derivative, $D$ denotes 'derivative' (RC stands for Riesz-Caputo), $a,t,b$ are so called terminals,  and the factor $\frac{\Gamma(2-\alpha)}{\Gamma(2)}$ will be clarified in Section \ref{ssec:SAO} where it appears for objectivity reasons. The terminals $a$ and $b$ can be chosen arbitrarily (in next section we will define the relation between the choice of terminals with the physical length scale of a particular material).

In Eq.~(\ref{eq:RC}) $_{a}^{C}D^{\alpha}_t f(t)$, $~_{t}^{C}D^{\alpha}_b f(t)$ are left and right Caputo's fractional derivatives, respectively. Their definitions are obtained from generalisations of the n-fold integration, so: 
\newline
left-sided Caputo's derivative $\textrm{for} \quad t>a$ and $n=[\alpha]+1$ is
\begin{equation}\label{eq:CaL}
~_{a}^{C}D^{\alpha}_t f(t)=\frac{1}{\Gamma(n-\alpha)}\int_a^t \frac{f^{(n)}(\tau)}{(t-\tau)^{\alpha-n+1}}d\tau;
\end{equation}
and right-sided Caputo derivative $\textrm{for} \quad t<b$ and $n=[\alpha]+1$ is
\begin{equation}\label{eq:CaR}
~_{t}^{C}D^{\alpha}_b f(t)=\frac{(-1)^n}{\Gamma(n-\alpha)}\int_t^b \frac{f^{(n)}(\tau)}{(\tau-t)^{\alpha-n+1}}d\tau.
\end{equation}
Notice, that both definitions include integration over the interval $(a,t)$ or $(t,b)$, respectively. It is clear for $\alpha=n\in \mathbb{N} \backslash \{0\}$ (then $m=\alpha$) classical derivative is captured and for $\alpha=0$ we have $~_{a}^{C}D^{0}_t f(t)=f(t)$. It should be emphasised, that the specific for Caputo's derivative is that for a constant function is equal zero and requires standard (like in the classical differential equations) initial and$/$or boundary conditions. 

In the remaining part of this paper the RC derivative is shortly denoted as $D^{\alpha}$ with the possibility of writing variable under the $D$ in case of partial differentiation of multivariate functions. For example $\underset{X_1}{D} ~^\alpha f $ represents partial fractional derivative of $f$ with respect to the variable $X_1$ over the interval which should be explicitly defined before $X_1\in(a,b)$. 
It is important that for $\alpha=1$ we have
\begin{equation}
~_{~~a}^{RC}D^{1}_b f(t)=\frac{\mathrm{d}}{\mathrm{d}t}f(t).
\end{equation}

\subsection{Fractional deformation gradients}\label{ssec:SAO}
The description is given in the Euclidean space. We refer to $\mathcal{B}$ as the reference configuration of the continuum body while $\mathcal{S}$ denotes its current configuration. Points in $\mathcal{B}$ are denoted by $\mathbf{X}$ and in $\mathcal{S}$ by $\mathbf{x}$. Coordinate system for $\mathcal{B}$ is denoted by $\left\{X_{A}\right \}$ with base $\mathbf{E}_{A}$ and for $\mathcal{S}$ we have $\left\{x_{a}\right \}$ with base $\mathbf{e}_{a}$.

We generalise the classical deformation gradient and its inverse as follows \cite{Sumelka2013-work2013,Sumelka2013-CMM2013}
\begin{equation}\label{eq:fracdefGradX}
\underset{X}{\tilde{\mathbf{F}}}(\mathbf{X},t)=\ell_X^{\alpha-1}\underset{X}{D}^\alpha\phi(\mathbf{X},t),\quad \mathrm{or}\quad {\underset{X}{\tilde{F}}} ~_{aA}=\ell_{A}^{\alpha-1}\underset{X_A}{D}^\alpha\phi_a\mathbf{e}_a\varotimes\mathbf{E}_A,
\end{equation}
and
\begin{equation}\label{eq:fracdefGradx}
\underset{x}{\tilde{\mathbf{F}}}(\mathbf{x},t)=\ell_x^{\alpha-1}\underset{x}{D}^\alpha\varphi(\mathbf{x},t),\quad \mathrm{or}\quad {\underset{x}{\tilde{F}}} ~_{Aa}=\ell_a^{\alpha-1}\underset{x_a}{D}^\alpha\varphi_A\mathbf{E}_A\varotimes\mathbf{e}_a,
\end{equation}
where $\underset{X}{\tilde{\mathbf{F}}}$ and $\underset{x}{\tilde{\mathbf{F}}}$ are fractional deformation gradients, ${D}^\alpha$ is a fractional differential operator in the sense of RC defined in previous section, and $\ell_X$ and $\ell_x$ are length scales in $\mathcal{B}$ and $\mathcal{S}$, respectively, $\phi$ defines the regular motion of the material body while $\varphi$ its inverse. We assume additionally that $\ell=\ell_X=\ell_x$, hence the isotropic non-locality is considered. It should be pointed out that the length scale must not be constant, thus can focus precisely experimental observations (one can model different length scales for elastic and plastic ranges) \cite{Voyiadjis2013-IJP}.

In general we have the following relations

\begin{equation}\label{eq:fracdefGradX}
\underset{x}{\tilde{\mathbf{F}}}\underset{X}{\tilde{\mathbf{F}}}\neq \mathbf{I}=\delta_{AB}\mathbf{E}_A\varotimes\mathbf{E}_B,
\end{equation}

and 

\begin{equation}\label{eq:fracdefGradX}
\underset{X}{\tilde{\mathbf{F}}}\underset{x}{\tilde{\mathbf{F}}}\neq \mathbf{i}=\delta_{ab}\mathbf{e}_a\varotimes\mathbf{e}_b,
\end{equation}
where $\delta$ denotes the Kronecker delta. It should be emphasised that fractional deformation gradients  $\underset{X}{\tilde{\mathbf{F}}}$ and $\underset{x}{\tilde{\mathbf{F}}}$  are non-local due to the definition of RC fractional differential operator which bases on an interval defined dependently on material being described.

Now, we can be introduce the following relations
\begin{equation}\label{eq:dxa}
\mathrm{d}\tilde{\mathbf{x}}=\underset{X}{\tilde{\mathbf{F}}}\mathrm{d}\mathbf{X},\quad \mathrm{or}\quad \mathrm{d}\tilde{x}_a=\underset{X}{\tilde{F}} ~_{aA}\mathrm{d}X_A\mathbf{e}_a,
\end{equation}
and 
\begin{equation}\label{eq:dXa}
\mathrm{d}\tilde{\mathbf{X}}=\underset{x}{\tilde{\mathbf{F}}}\mathrm{d}\mathbf{x},\quad \mathrm{or}\quad \mathrm{d}\tilde{X}_A=\underset{x}{\tilde{F}} ~_{Aa}\mathrm{d}x_a\mathbf{E}_A,
\end{equation}

where $\mathrm{d}\tilde{\mathbf{x}}$ and $\mathrm{d}\tilde{\mathbf{X}}$ are the fractional spatial and material line elements, respectively, while $\mathrm{d}\mathbf{x}$ and $\mathrm{d}\mathbf{X}$ are classical spatial and material line elements, respectively.

\begin{figure}[H]
\centering
\includegraphics[width=10cm]{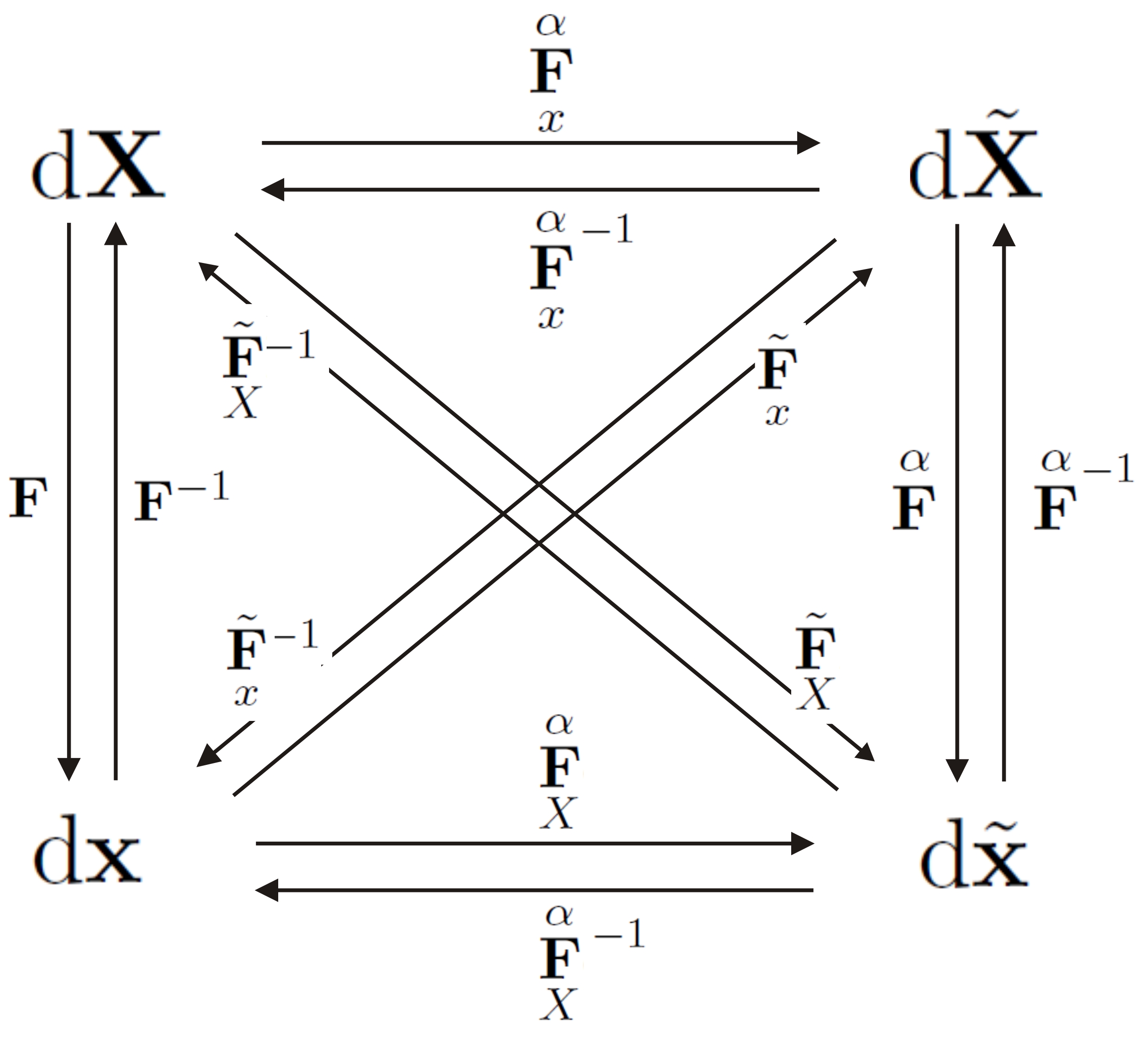}
\caption{The relations between material and spatial line elements with their fractional counterparts}
\label{fig:diagram}
\end{figure}

We have additionally (cf. Fig.~\ref{fig:diagram}):
\begin{equation}\label{eq:dxaII}
\mathrm{d}\tilde{\mathbf{x}}=\overset{\alpha}{\mathbf{F}}\mathrm{d}\tilde{\mathbf{X}},\quad \mathrm{or}\quad \mathrm{d}\tilde{x}_a=\overset{\alpha}{F}_{aA}\mathrm{d}\tilde{X}_A\mathbf{e}_a,
\end{equation}

\begin{equation}\label{eq:dXXa}
\mathrm{d}\tilde{\mathbf{X}}=\underset{x}{\overset{\alpha}{\mathbf{F}}}\mathrm{d}\mathbf{X},\quad \mathrm{or}\quad \mathrm{d}\tilde{X}_B=\underset{x}{\overset{\alpha}{{F}}} ~_{BA}\mathrm{d}X_A\mathbf{E}_B,
\end{equation}

\begin{equation}\label{eq:dXXa}
\mathrm{d}\tilde{\mathbf{x}}=\underset{X}{\overset{\alpha}{\mathbf{F}}}\mathrm{d}\mathbf{x},\quad \mathrm{or}\quad \mathrm{d}\tilde{x}_b=\underset{X}{\overset{\alpha}{F}} ~_{ba}\mathrm{d}x_a\mathbf{e}_b,
\end{equation}

where $\overset{\alpha}{\mathbf{F}}=\underset{X}{\tilde{\mathbf{F}}}\mathbf{F}^{-1}\underset{x}{\tilde{\mathbf{F}}} ~^{-1}$, $\underset{x}{\overset{\alpha}{\mathbf{F}}}=\underset{x}{\tilde{\mathbf{F}}}\mathbf{F}$ and $\underset{X}{\overset{\alpha}{\mathbf{F}}}=\underset{X}{\tilde{\mathbf{F}}}\mathbf{F}^{-1}$. 
It is clear that $\underset{x}{\overset{\alpha}{\mathbf{F}}}$ and $\underset{X}{\overset{\alpha}{\mathbf{F}}}$ are not two point tensors while $\underset{X}{\tilde{\mathbf{F}}}$, $\underset{x}{\tilde{\mathbf{F}}}$ and $\overset{\alpha}{\mathbf{F}}$ are. Based on the properties of motion the inverse of $\underset{X}{\tilde{\mathbf{F}}}$ and $\underset{x}{\tilde{\mathbf{F}}}$ exists.

It is important, that to fulfil objectivity requirements the length scale $\ell$ is not arbitrary. As mentioned the length scale is closely related to the choice of terminals in fractional differential operator. It can be thought as  putting the physical constraints on the obtained fractional generalisation of classical kinematics.
The described situation is clear when one calculates $\underset{X}{\tilde{\mathbf{F}}}$ and $\underset{x}{\tilde{\mathbf{F}}}$ for the rigid-body motion under the assumption that in RC fractional derivative the terminals are $a=X_A-\frac{L}{2}$ and $b=\frac{L}{2}+X_A$ (thus we calculate the RC derivative on the interval with length L, and $X_A$ is the point of interest). Hence we have
\begin{equation}\label{eq:1}
\underset{X}{\tilde{\mathbf{F}}}=\ell^{\alpha-1}\left(\frac{L}{2}\right)^{1-\alpha}\mathbf{R},
\end{equation}
and
\begin{equation}\label{eq:2}
\underset{x}{\tilde{\mathbf{F}}}=\ell^{\alpha-1}\left(\frac{L}{2}\right)^{1-\alpha}\mathbf{R}^{-1},
\end{equation}
where $\mathbf{R}$ is an orthogonal tensor.
Thus, from Eqs~(\ref{eq:1}) and (\ref{eq:2}) we see that only for 
\begin{equation}\label{eq:ellDefin}
\ell=\frac{L}{2},
\end{equation}
the pure rotation is obtained. This relation is chosen as a definition of $\ell$.

Knowing the explicit definition of $\ell$ the transformations of deformation gradients under superimposed rigid-body motion can be expressed as:
\begin{equation}\label{eq:aa}
\mathbf{F}^*=\mathbf{Q}\mathbf{F},
\end{equation}

\begin{equation}
\underset{X}{\tilde{\mathbf{F}}}^*=\mathbf{Q}\underset{X}{\tilde{\mathbf{F}}},
\end{equation}

\begin{equation}
\underset{x}{\tilde{\mathbf{F}}}^*=\underset{x}{\tilde{\mathbf{F}}} \mathbf{Q}^{-1},
\end{equation}
so
\begin{equation}
{\overset{\alpha}{\mathbf{F}}} ~^*=\mathbf{Q}\underset{X}{\tilde{\mathbf{F}}}\mathbf{F}^{-1}\mathbf{Q}^{-1}(\underset{x}{\tilde{\mathbf{F}}} \mathbf{Q}^{-1})^{-1}=\mathbf{Q}\underset{X}{\tilde{\mathbf{F}}}\mathbf{F}^{-1}\underset{x}{\tilde{\mathbf{F}}}=\mathbf{Q}{\overset{\alpha}{\mathbf{F}}}, 
\end{equation}

\begin{equation}
\underset{X}{\overset{\alpha}{\mathbf{F}}} ~^*= \mathbf{Q}\underset{X}{\tilde{\mathbf{F}}}\mathbf{F}^{-1}\mathbf{Q}^{-1}=\mathbf{Q}\underset{X}{\overset{\alpha}{\mathbf{F}}}\mathbf{Q}^{T},
 \end{equation}

\begin{equation}\label{eq:bb}
\underset{x}{\overset{\alpha}{\mathbf{F}}} ~^*= \underset{x}{\tilde{\mathbf{F}}}\mathbf{Q}^{-1}\mathbf{Q}\mathbf{F}=\underset{x}{\overset{\alpha}{\mathbf{F}}},
 \end{equation}
where $\mathbf{Q}(t)$ is assumed to be the proper orthogonal tensor, and $(\cdot)~^*$ denotes new coordinate system. Thus, from Eqs~(\ref{eq:aa})-(\ref{eq:bb}) it appears that all fractional counterparts of classical measures of deformation keep the same objectivity relations. Similarly to the classical approach also in this approach any material field is unaffected by a rigid-body motion superimposed on current configuration.

\begin{figure}[H]
\centering
\includegraphics[width=15cm]{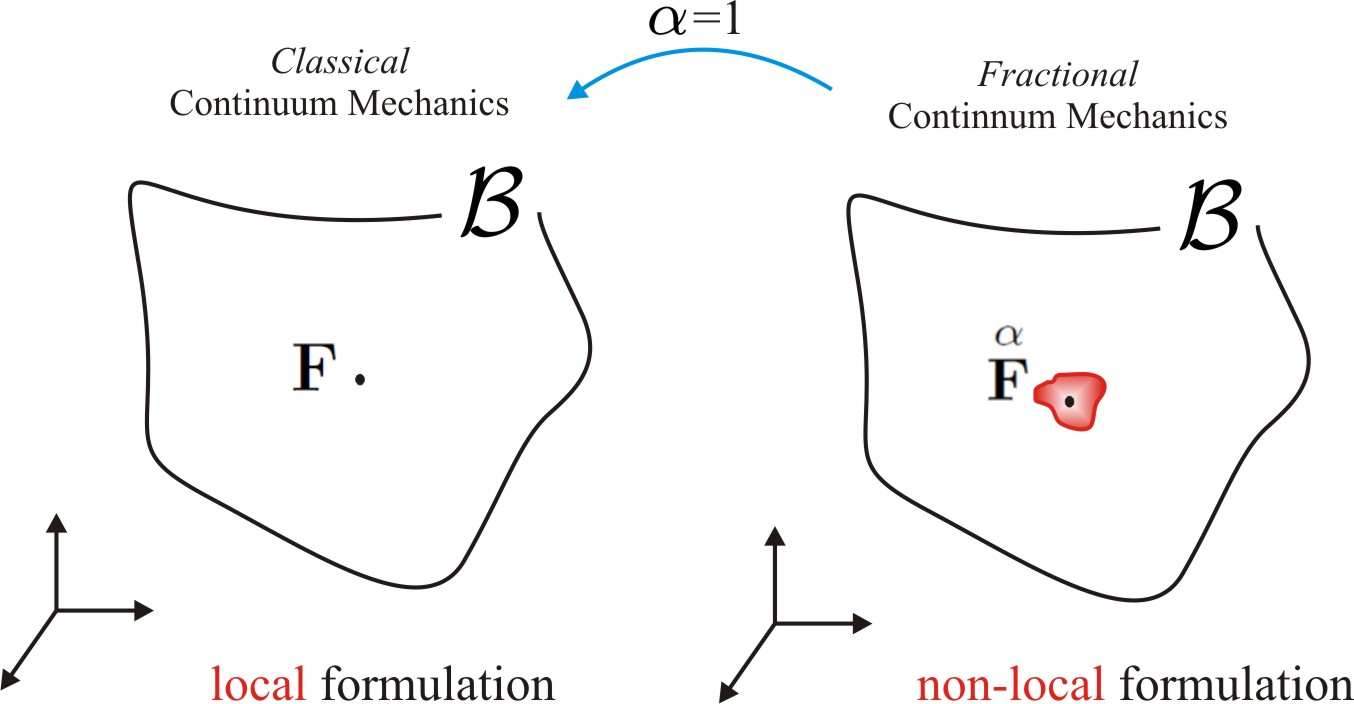}
\caption{The equivalence of fractional continua and classical continua for $\alpha=1$}
\label{fig:frac-class}
\end{figure}

Finally, taking $\alpha=1$ (RC fractional derivative becomes classical derivative) we recover classical local continuum mechanics (where $\ell^{\alpha-1}=\ell^{1-1}=\ell^{0}=1$ does not influence the results), so (Fig.~\ref{fig:frac-class}):
\begin{equation}
\mathbf{F}=\underset{X}{\tilde{\mathbf{F}}}=\underset{x}{\tilde{\mathbf{F}}} ~^{-1}={\overset{\alpha}{\mathbf{F}}},
\end{equation}

\begin{equation}
\mathbf{F}^{-1}=\underset{X}{\tilde{\mathbf{F}}} ~^{-1}=\underset{x}{\tilde{\mathbf{F}}}={\overset{\alpha}{\mathbf{F}}} ~^{-1},
\end{equation}

\begin{equation}
\underset{x}{\overset{\alpha}{\mathbf{F}}}=\mathbf{I},
\end{equation}

\begin{equation}
\underset{X}{\overset{\alpha}{\mathbf{F}}}=\mathbf{i},
\end{equation}

\begin{equation}
\mathrm{d}{\mathbf{x}}=\mathrm{d}\tilde{\mathbf{x}},
\end{equation}

\begin{equation}
\mathrm{d}{\mathbf{X}}=\mathrm{d}\tilde{\mathbf{X}}.
\end{equation}

\subsection{Total fractional strains}\label{sse:frstr}
We define the total fractional strains by analogy to the classical continuum mechanics. Thus, the difference in scalar products in actual and reference configurations allows to define 4 concepts of finite strains:

\begin{enumerate}
    \item Classical formulation
		\begin{equation}
		\mathrm{d}{\mathbf{x}}\mathrm{d}{\mathbf{x}}-
\mathrm{d}{\mathbf{X}}\mathrm{d}{\mathbf{X}}\equiv
\mathrm{d}{\mathbf{X}}(\mathbf{F}^T\mathbf{F}-\mathbf{I})\mathrm{d}{\mathbf{X}}\equiv
\mathrm{d}{\mathbf{x}}(\mathbf{i}-\mathbf{F}^{-T}\mathbf{F}^{-1})\mathrm{d}{\mathbf{x}},
		\end{equation}
so
\begin{equation}
		\mathrm{E}=\frac{1}{2}(\mathbf{F}^T\mathbf{F}-\mathbf{I}), \quad
\mathbf{e}=\frac{1}{2}(\mathbf{i}-\mathbf{F}^{-T}\mathbf{F}^{-1}).
		\end{equation}
	\item Formulation based on the fractional spatial line element ($\mathrm{d}{\tilde{\mathbf{x}}}$) and the classical material line element $(\mathrm{d}{\mathbf{X}})$

		\begin{equation}
		\mathrm{d}{\tilde{\mathbf{x}}}\mathrm{d}{\tilde{\mathbf{x}}}-
\mathrm{d}{\mathbf{X}}\mathrm{d}{\mathbf{X}}\equiv
\mathrm{d}{\mathbf{X}}(\underset{X}{\tilde{\mathbf{F}}} ~^T\underset{X}{\tilde{\mathbf{F}}}-\mathbf{I})\mathrm{d}{\mathbf{X}}\equiv
\mathrm{d}{\tilde{\mathbf{x}}}(\mathbf{i}-\underset{X}{\tilde{\mathbf{F}}} ~^{-T}\underset{X}{\tilde{\mathbf{F}}} ~^{-1})\mathrm{d}{\tilde{\mathbf{x}}},
		\end{equation}
so
\begin{equation}
		\underset{X}{\tilde{\mathbf{E}}}=\frac{1}{2}(\underset{X}{\tilde{\mathbf{F}}} ~^T\underset{X}{\tilde{\mathbf{F}}}-\mathbf{I}), \quad
\underset{X}{\tilde{\mathbf{e}}}=\frac{1}{2}(\mathbf{i}-\underset{X}{\tilde{\mathbf{F}}} ~^{-T}\underset{X}{\tilde{\mathbf{F}}} ~^{-1}).
		\end{equation}

	\item Formulation based on the classical spatial line element ($\mathrm{d}{{\mathbf{x}}}$) and the fractional material line element $(\mathrm{d}{\tilde{\mathbf{X}}})$

		\begin{equation}
		\mathrm{d}{\mathbf{x}}\mathrm{d}{\mathbf{x}}-
\mathrm{d}{\tilde{\mathbf{X}}}\mathrm{d}{\tilde{\mathbf{X}}}\equiv
\mathrm{d}{\tilde{\mathbf{X}}}(\underset{x}{\tilde{\mathbf{F}}} ~^{-T}\underset{x}{\tilde{\mathbf{F}}} ~^{-1}-\mathbf{I})\mathrm{d}{\tilde{\mathbf{X}}}\equiv
\mathrm{d}{{\mathbf{x}}}(\mathbf{i}-\underset{x}{\tilde{\mathbf{F}}} ~^{T}\underset{x}{\tilde{\mathbf{F}}})\mathrm{d}{{\mathbf{x}}},
		\end{equation}
so
\begin{equation}
		\underset{x}{\tilde{\mathbf{E}}}=\frac{1}{2}(\underset{x}{\tilde{\mathbf{F}}} ~^{-T}\underset{x}{\tilde{\mathbf{F}}} ~^{-1}-\mathbf{I}), \quad
\underset{x}{\tilde{\mathbf{e}}}=\frac{1}{2}(\mathbf{i}-\underset{x}{\tilde{\mathbf{F}}} ~^{T}\underset{x}{\tilde{\mathbf{F}}} ).
		\end{equation}

	\item Formulation based on the fractional spatial line element ($\mathrm{d}{\tilde{\mathbf{x}}}$) and fractional material line element $(\mathrm{d}{\tilde{\mathbf{X}}})$

		\begin{equation}
		\mathrm{d}{\tilde{\mathbf{x}}}\mathrm{d}{\tilde{\mathbf{x}}}-
\mathrm{d}{\tilde{\mathbf{X}}}\mathrm{d}{\tilde{\mathbf{X}}}\equiv
\mathrm{d}{\tilde{\mathbf{X}}}(\overset{\alpha}{\mathbf{F}} ~^T\overset{\alpha}{\mathbf{F}}-\mathbf{I})\mathrm{d}{\tilde{\mathbf{X}}}\equiv
\mathrm{d}{\tilde{\mathbf{x}}}(\mathbf{i}-\overset{\alpha}{\mathbf{F}} ~^{-T}\overset{\alpha}{\mathbf{F}} ~^{-1})\mathrm{d}{\tilde{\mathbf{x}}},
		\end{equation}
so
\begin{equation}
		\overset{\alpha}{\mathbf{E}}=\frac{1}{2}(\overset{\alpha}{\mathbf{F}} ~^T{\overset{\alpha}{\mathbf{F}}}-\mathbf{I}), \quad
\overset{\alpha}{\mathbf{e}}=\frac{1}{2}(\mathbf{i}-\overset{\alpha}{\mathbf{F}} ~^{-T}\overset{\alpha}{\mathbf{F}} ~^{-1}).
		\end{equation}
	
\end{enumerate}   

Taking into account the results of previous section it is clear that the generalisation of classical continuum mechanics can be formulated just by exchanging classical deformation gradient with the one of its fractional counterparts. Thus, one can define:  

		\begin{equation}\label{eq:Efrac}
			\mathbf{E}=\frac{1}{2}(\overset{\Diamond}{\mathbf{F}} ~^T\overset{\Diamond}{\mathbf{F}}-\mathbf{I}),\quad \mathrm{or}\quad {E}_{AB}=\frac{1}{2}(\overset{\Diamond}{F} ~^T_{Aa}\overset{\Diamond}{F} ~_{aB}-{I}_{AB})\mathbf{E}_A\varotimes\mathbf{E}_B,
\end{equation}

		\begin{equation}\label{eq:efrac}			
		\mathbf{e}=\frac{1}{2}(\mathbf{i}-\overset{\Diamond}{\mathbf{F}} ~^{-T}\overset{\Diamond}{\mathbf{F}} ~^{-1}),\quad \mathrm{or}\quad {e}_{ab}=\frac{1}{2}({i}_{ab}-\overset{\Diamond}{F} ~_{aA}^{-T}\overset{\Diamond}{F} ~^{-1}_{Ab})\mathbf{e}_a\varotimes\mathbf{e}_b,
		\end{equation}

where generalised pull-back transformation of $\mathbf{e}$ gives
		\begin{equation}
			\mathbf{E}=\overset{\Diamond}{\mathbf{F}} ~^T\mathbf{e}\overset{\Diamond}{\mathbf{F}},
\end{equation}

while generalised push-forward of $\mathbf{E}$ gives
		\begin{equation}
			\mathbf{e}=\overset{\Diamond}{\mathbf{F}} ~^{-T}\mathbf{E}\overset{\Diamond}{\mathbf{F}} ~^{-1},
\end{equation}

and
		\begin{equation}
			\mathbf{C}=\overset{\Diamond}{\mathbf{F}} ~^T\overset{\Diamond}{\mathbf{F}},\quad \mathrm{or}\quad {C}_{AB}=\overset{\Diamond}{F} ~_{Aa}^T\overset{\Diamond}{F} ~_{aB}\mathbf{E}_A\varotimes\mathbf{E}_B,
		\end{equation}

		\begin{equation}			
		\mathbf{c}=\overset{\Diamond}{\mathbf{F}} ~^{-T}\overset{\Diamond}{\mathbf{F}} ~^{-1}=\mathbf{b}^{-1},\quad \mathrm{or}\quad {c}_{ab}=\overset{\Diamond}{F} ~_{aA}^{-T}\overset{\Diamond}{F} ~^{-1}_{Ab}\mathbf{e}_a\varotimes\mathbf{e}_b,
		\end{equation}

finally using the theorem of polar decomposition of non-singular second order tensor we have

		\begin{equation}			
		\overset{\Diamond}{\mathbf{F}}=\mathbf{R}\mathbf{U}=\mathbf{v}\mathbf{R},\quad \mathrm{or}\quad \overset{\Diamond}{F} ~_{aA}=R_{aB}U_{BA}\mathbf{e}_a\varotimes\mathbf{E}_B=v_{ab}R_{bA}\mathbf{e}_a\varotimes\mathbf{E}_B,
		\end{equation}

and as a consequence

		\begin{equation}			
\mathbf{C}=\mathbf{U}\mathbf{U}\quad \mathrm{and}\quad \mathbf{b}=\mathbf{v}\mathbf{v}.
		\end{equation}

In the above expressions, depending on the formulation, $\overset{\Diamond}{\mathbf{F}}$ can be replaced with $\mathbf{F}$ or $\underset{X}{\tilde{\mathbf{F}}}$ or $\underset{x}{\tilde{\mathbf{F}}}$ or $\overset{\alpha}{\mathbf{F}}$.  
According to the chosen $\overset{\Diamond}{\mathbf{F}}$ the associated others variables denote: $\mathbf{E}$ is the classical Green-Lagrange strain tensor or its fractional counterpart (symmetric); $\mathbf{e}$ is the classical Euler-Almansi strain tensor or its fractional counterpart (symmetric); $\mathbf{C}$ is the classical right Cauchy-Green tensor or its fractional counterpart (symmetric and positive definite); $\mathbf{c}=\mathbf{b}^{-1}$ is the classical left Cauchy-Green tensor/Finger deformation tensor or its fractional counterpart (symmetric and positive definite); $\mathbf{R}$ is orthogonal tensor; $\mathbf{U}$ is the classical right stretch tensor (symmetric and positive definite) or its fractional counterpart, $\mathbf{v}$ is the classical left stretch tensor (symmetric and positive definite) or its fractional counterpart.

\begin{figure}[H]
\centering
\includegraphics[width=10cm]{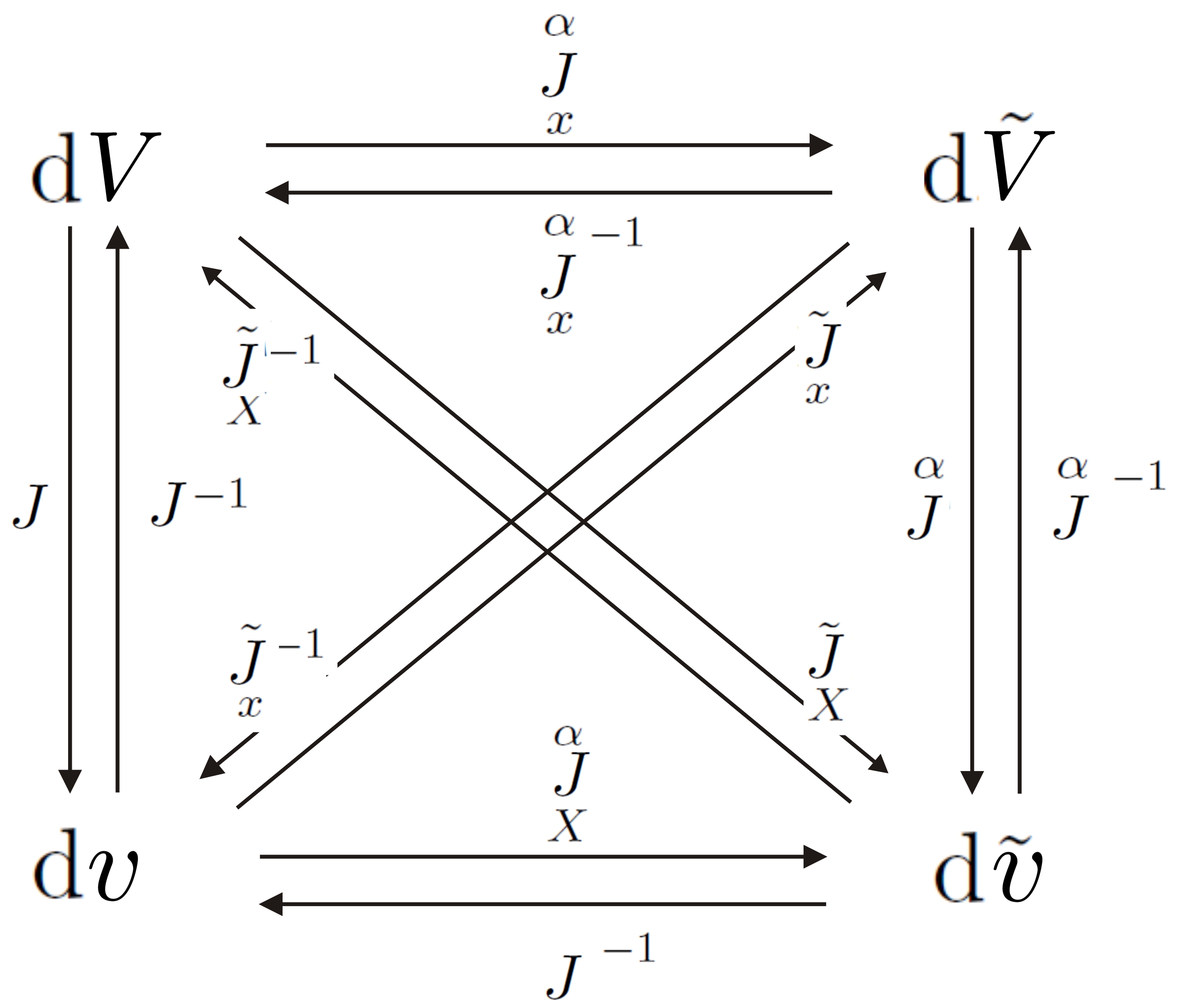}
\caption{The relations between material and spatial volumes with their fractional counterparts}
\label{fig:diagram-V}
\end{figure}

\begin{figure}[H]
\centering
\includegraphics[width=10cm]{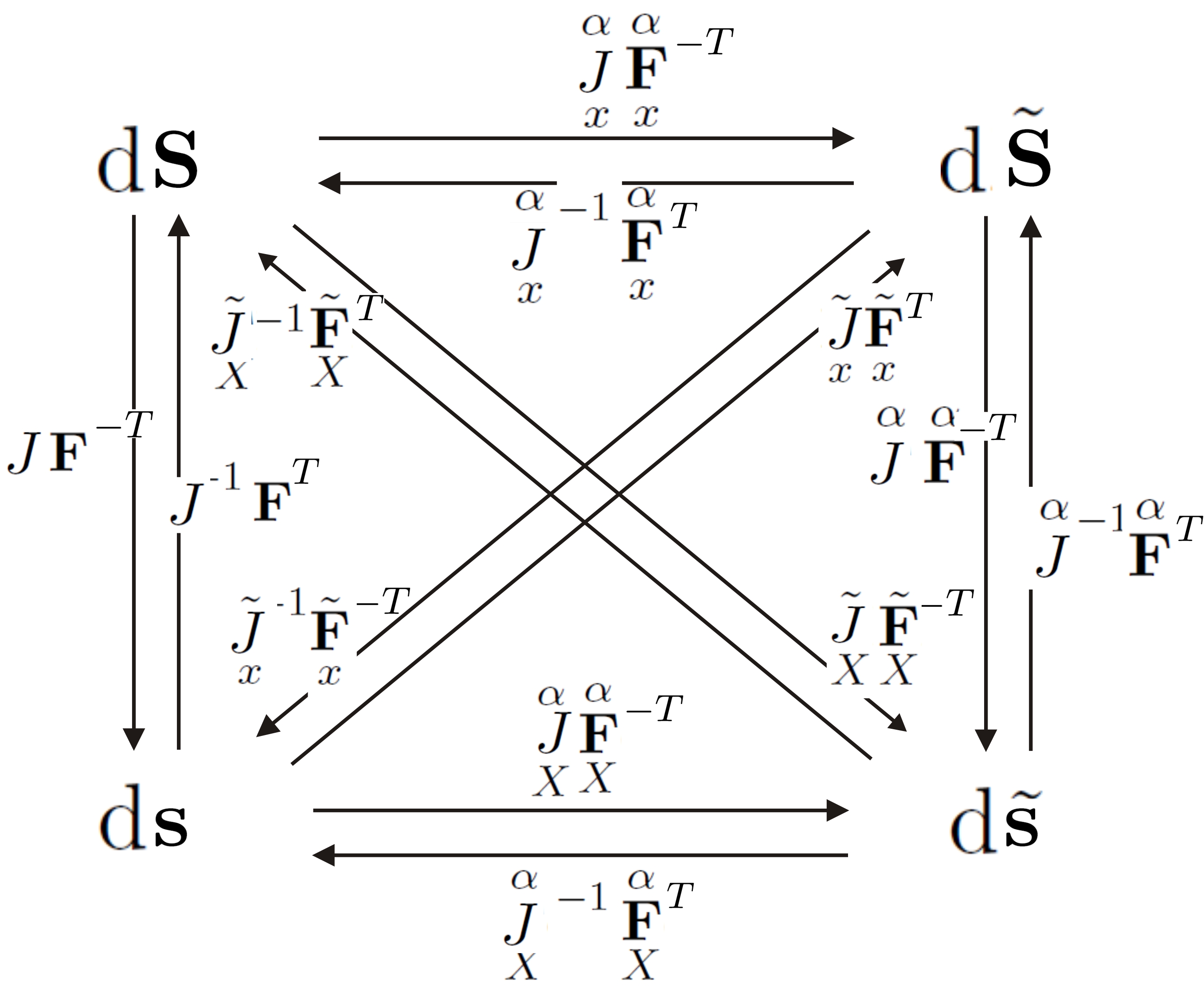}
\caption{The relations between material and spatial vector elements with their fractional counterparts}
\label{fig:diagram-S}
\end{figure}

We have also analogous definitions for the volume ratio and surface element mapping, we have
\begin{equation}
\mathrm{d}v=\det (\overset{\Diamond}{\mathbf{F}}) \mathrm{d}V,
 \end{equation}
and
\begin{equation}
\mathrm{d}\mathbf{s}=\det (\overset{\Diamond}{\mathbf{F}}) \overset{\Diamond}{\mathbf{F}} ~^{-T} \mathrm{d}\mathbf{S},
 \end{equation}
where according to chosen $\overset{\Diamond}{\mathbf{F}}$ the following variables denote: $\mathrm{d}v$ is the spatial volume element or its fractional counterpart, $\mathrm{d}V$ is the material volume element or its fractional counterpart, $\mathrm{d}\mathbf{s}$ is the spatial vector element or its fractional counterpart, $\mathrm{d}\mathbf{S}$ is the material vector element or its fractional counterpart - cf. Figs~\ref{fig:diagram-V} and \ref{fig:diagram-S} (by analogy to Fig.~\ref{fig:diagram}). In Figs~\ref{fig:diagram-V} and \ref{fig:diagram-S} we have denoted: $J=\det(\mathbf{F})$, $\underset{X}{\tilde{J}}=\det(\underset{X}{\tilde{\mathbf{F}}})$, $\underset{x}{\tilde{J}}=\det(\underset{x}{\tilde{\mathbf{F}}})$, $\underset{X}{\overset{\alpha}{J}}=\det(\underset{X}{\overset{\alpha}{\mathbf{F}}})$, 
$\underset{x}{\overset{\alpha}{J}}=\det(\underset{x}{\overset{\alpha}{\mathbf{F}}})$,
${\overset{\alpha}{J}}=\det({\overset{\alpha}{\mathbf{F}}})$.

\subsection{Infinitesimal total fractional strains}

As in the classical continuum mechanics one can define the relation between fractional strains and fractional displacement gradient tensor utilising introduced fractional gradient tensors $\underset{X}{\tilde{\mathbf{F}}}$ and $\underset{x}{\tilde{\mathbf{F}}}$. Then, after omitting the higher order terms we obtain small total fractional strains.

The displacements in the material description $\mathbf{U}$ are defined as ($\mathbf{U}$ and should not be confused with the right stretch tensor defined previously):
\begin{equation}
\mathbf{U}(\mathbf{X},t)=\mathbf{x}(\mathbf{X},t)-\mathbf{X},
\end{equation}
and its fractional gradient
\begin{equation}
\mathrm{Grad}\underset{X}{\tilde{\mathbf{U}}}=\underset{X}{\tilde{\mathbf{F}}}-\mathbf{I} ,\quad \mathrm{or}\quad \ell^{\alpha-1}\underset{X_A}{D}^\alpha U_a=(\underset{X}{\tilde{{F}}} ~_{aA} - I_{aA})\mathbf{e}_a\varotimes\mathbf{E}_A,
\end{equation}
thus we have
\begin{equation}\label{eq:fracdefXinU}
\underset{X}{\tilde{\mathbf{F}}}=\mathrm{Grad}\underset{X}{\tilde{\mathbf{U}}}+\mathbf{I}.
\end{equation}

Similarly, the displacements in spatial description $\mathbf{u}$ are defined as
\begin{equation}
\mathbf{u}(\mathbf{x},t)=\mathbf{x}-\mathbf{X}(\mathbf{x},t),
\end{equation}
and its fractional gradient
\begin{equation}
\mathrm{grad}\underset{x}{\tilde{\mathbf{u}}}=\mathbf{i}-\underset{x}{\tilde{\mathbf{F}}},\quad \mathrm{or}\quad \ell^{\alpha-1}\underset{x_a}{D}^\alpha u_A=(i_{Aa}-\underset{x}{\tilde{{F}}} ~_{Aa})\mathbf{E}_A\varotimes\mathbf{e}_a,
\end{equation}
thus we have
\begin{equation}\label{eq:fracdefxinu}
\underset{x}{\tilde{\mathbf{F}}}=\mathbf{i}-\mathrm{grad}\underset{x}{\tilde{\mathbf{u}}}.
\end{equation}

By applying Eqs~(\ref{eq:fracdefXinU}) and (\ref{eq:fracdefxinu}) into the fractional strain definitions Eqs~(\ref{eq:Efrac}) and (\ref{eq:efrac}) we obtain their dependence on the fractional displacement gradients. Thus one obtains:

\begin{equation}
\underset{X}{\tilde{\mathbf{E}}}=\frac{1}{2}(\mathrm{Grad}\underset{X}{\tilde{\mathbf{U}}}+\mathrm{Grad}\underset{X}{\tilde{\mathbf{U}}}^T
+\mathrm{Grad}\underset{X}{\tilde{\mathbf{U}}}^T\mathrm{Grad}\underset{X}{\tilde{\mathbf{U}}}),
\end{equation}

\begin{equation}
\underset{X}{\tilde{\mathbf{e}}}=\frac{1}{2}(-\mathrm{Grad}\underset{X}{\tilde{\mathbf{U}}}^{-1}-
\mathrm{Grad}\underset{X}{\tilde{\mathbf{U}}}^{-T}
-\mathrm{Grad}\underset{X}{\tilde{\mathbf{U}}}^{-T}\mathrm{Grad}\underset{X}{\tilde{\mathbf{U}}}^{-1}),
\end{equation}
 and
\begin{equation}
		\underset{x}{\tilde{\mathbf{E}}}=\frac{1}{2}(-\mathrm{grad}\underset{x}{\tilde{\mathbf{u}}}^{-1}-
\mathrm{grad}\underset{x}{\tilde{\mathbf{u}}}^{-T}
+\mathrm{grad}\underset{x}{\tilde{\mathbf{u}}}^{-T}\mathrm{grad}\underset{x}{\tilde{\mathbf{u}}^{-1}}),
\end{equation}

\begin{equation}
\underset{x}{\tilde{\mathbf{e}}}=\frac{1}{2}(\mathrm{grad}\underset{x}{\tilde{\mathbf{u}}}+
\mathrm{grad}\underset{x}{\tilde{\mathbf{u}}}^{T}
-\mathrm{grad}\underset{x}{\tilde{\mathbf{u}}}^{T}\mathrm{grad}\underset{x}{\tilde{\mathbf{u}}}),
\end{equation}
and

\begin{equation}
		\overset{\alpha}{\mathbf{E}}=\frac{1}{2}\left[(\mathrm{Grad}\underset{X}{\tilde{\mathbf{U}}}+\mathrm{Grad}\underset{X}{\tilde{\mathbf{U}}}^T)
-(\mathrm{grad}\underset{x}{\tilde{\mathbf{u}}}^{-1}+\mathrm{grad}\underset{x}{\tilde{\mathbf{u}}}^{-T})
-(\nabla\mathbf{u}+\nabla\mathbf{u}^T)+(...)\right], 
\end{equation}

\begin{equation}
\overset{\alpha}{\mathbf{e}}=\frac{1}{2}\left[-(\mathrm{Grad}\underset{X}{\tilde{\mathbf{U}}}^{-1}+\mathrm{Grad}\underset{X}{\tilde{\mathbf{U}}}^{-T})
+(\mathrm{grad}\underset{x}{\tilde{\mathbf{u}}}+\mathrm{grad}\underset{x}{\tilde{\mathbf{u}}}^{T})
+(\nabla\mathbf{u}^{-1}+\nabla\mathbf{u}^{-T})+(...)\right].
\end{equation}

For definitions $\overset{\alpha}{\mathbf{E}}$ and $\overset{\alpha}{\mathbf{e}}$ we have omitted second and third ordered terms denoting them $(...)$ for clarity.

Of course for $\alpha=1$ we have classical solution ($\mathrm{Grad}\underset{X}{\tilde{\mathbf{U}}}^{-1}=-\nabla\mathbf{u}$ and $\mathrm{grad}\underset{x}{\tilde{\mathbf{u}}}^{-1}=-\nabla\mathbf{U}$ like in classical continuum mechanics where $\nabla\mathbf{u}=-\nabla\mathbf{U}^{-1}$ and consequently  $\nabla\mathbf{U}=-\nabla\mathbf{u}^{-1}$), so

\begin{equation}
		\mathrm{E}=\frac{1}{2}(\nabla\mathbf{U}+\nabla\mathbf{U}^T+\nabla\mathbf{U}^T\nabla\mathbf{U})
=\frac{1}{2}(-\nabla\mathbf{u}^{-1}-\nabla\mathbf{u}^{-T}+\nabla\mathbf{u}^{-T}\nabla\mathbf{u}^{-1}),
		\end{equation}

\begin{equation}
\mathbf{e}=\frac{1}{2}(\nabla\mathbf{u}+\nabla\mathbf{u}^T-\nabla\mathbf{u}^T\nabla\mathbf{u})
=\frac{1}{2}(-\nabla\mathbf{U}^{-1}-\nabla\mathbf{U}^{-T}-\nabla\mathbf{U}^{-T}\nabla\mathbf{U}^{-1}),
		\end{equation}

where $\nabla$ stands for classical gradient.

If we now take into account small deformation assumption, understand as omitting higher order terms in above strain definitions, we obtain infinitesimal fractional total Cauchy strain tensor
\begin{equation}
\overset{\Diamond}{\bveps}=\frac{1}{2}\left[\mathrm{Grad}\underset{X}{\tilde{\mathbf{U}}}
+\mathrm{Grad}\underset{X}{\tilde{\mathbf{U}}}^T\right]=\frac{1}{2}\left[\mathrm{grad}\underset{x}{\tilde{\mathbf{u}}}
+\mathrm{grad}\underset{x}{\tilde{\mathbf{u}}}^T\right].
\end{equation}
And again for $\alpha=1$ classical Cauchy strain tensor is recovered
\begin{equation}
\bveps=\overset{\Diamond}{\bveps}=\frac{1}{2}\left[\nabla\mathbf{U}
+\nabla\mathbf{U}^T\right]=\frac{1}{2}\left[\nabla\mathbf{u}
+\nabla\mathbf{u}^T\right].
\end{equation}

\subsection{Physical interpretation}
The non-local fractional kinematics defines fractional continua. We add two material parameters $\alpha$ and $\ell$ in comparison with classical formulation. The first is the order of fractional continua $\alpha$ (together with type of applied fractional differential operator) which controls way in which the information from the surrounding influences particular point of interest. The second is the length scale $\ell$ which defines the amount of this information (size of this surrounding). Both parameters are crucial. 

It is important, that the role of the length scale parameter appearing in definitions Eqs~(\ref{eq:fracdefGradX}) and (\ref{eq:fracdefGradx}) is  two fold. Firstly, without $\ell$ the unit of the fractional deformation gradients would be $m^{1-\alpha}$, hence introduction of the length, similarly like in the classical non-local gradient methods, allows to finally obtain dimensionless quantity. In this way, we can compare the lengths of line elements $\mathrm{d}{\mathbf{X}}$ and $\mathrm{d}{\mathbf{x}}$ with their fractional counterparts $\mathrm{d}\tilde{\mathbf{X}}$ and $\mathrm{d}\tilde{\mathbf{x}}$ what would be crucial concerning possible strains definitions. Secondly, it is necessary to introduce the length scale in order to fulfil the rigid body motion requirements. Notice that from purely mathematical point of view, those parameters could be omitted, however it is not the case in physical theory.

Concluding, the physical interpretation of all measures defined thorough this paper (e.g. fractional deformation gradients or fractional strains) remains unchanged compared with classical one - the only difference is that they operate on fractional or mixed (classical/factional) line elements and are non-local.

Recall that for $\alpha=1$ classical local rate independent plasticity model is recovered.

\section{Non-local fractional model of rate independent plasticity}
\label{ssec:BVP}

Let us consider the problem of static deformation under the assumption that material is elasto-plastic and small fractional deformation holds. The additive decomposition of total fractional strains $\overset{\Diamond}{\bveps}$ into elastic and plastic parts is assumed. The classical associative flow rule is considered.

The governing equations for boundary value problem of non-local fractional model of rate independent plasticity are: 
\begin{equation}\label{eq:govEqn}
\begin{cases}
\sigma_{ij,j}+b_i=0,\\
\overset{\Diamond}{\varepsilon}_{ij}=\frac{1}{2}\ell^{\alpha-1}\left(\underset{X_j}{D}^\alpha U_i+\underset{X_i}{D}^\alpha U_j\right),\\
\overset{\Diamond}{\varepsilon}_{ij}=\varepsilon^e_{ij}+\varepsilon^p_{ij},\\
\sigma_{ij}=\mathcal{L}^e_{ijkl}\varepsilon^e_{kl}=\mathcal{L}^e_{ijkl}(\overset{\Diamond}{\varepsilon}_{kl}-\varepsilon^p_{kl}),\\
\dot{\varepsilon}^p_{ij}=\gamma \frac{\partial f}{\partial \sigma_{ij}},\\
f=\sqrt{\tau_{ij}\tau_{ij}}-\sigma_Y\sqrt{\frac{2}{3}},\\
\gamma\geq 0, \quad f(\sigma_{ij})\leq 0, \quad \gamma f(\sigma_{ij})=0,\\
\gamma\dot{f}(\sigma_{ij})=0,\\
U_i=\check{U}_i,\quad\quad \mathbf{X}\in \Gamma_U,\\
\sigma_{ij}n_j=\check{t}_i,\quad\quad \mathbf{X}\in \Gamma_\sigma,\\
\Gamma_U\cap\Gamma_\sigma=\emptyset\quad\mathrm{and}\quad\Gamma_U\cup\Gamma_\sigma=\Gamma.
\end{cases} 
\end{equation}
In above we have denoted: $\bsigma$ is the Cauchy stress tensor, $\mathbf{b}$ is the body force, $\mathcal{L}^e$ is the stiffness tensor, $\bveps^e$ is the elastic strain tensor, $\bveps^p$ is the plastic strain tensor, $\gamma$ is consistency parameter, $f$ is yield function, $\bta$ is stress deviator, $\sigma_Y$ is flow stress, $\mathbf{n}$ is the outward unit normal vector, $\mathbf{t}$ is the Cauchy traction vector, $\Gamma_U$ and $\Gamma_\sigma$ are parts of boundary $\Gamma$ where the displacements and the tractions are applied, respectively.

\section{Numerical Examples}

\subsection{General remarks}
We focus the attention to the geometry of plastic strains zone and their magnitude. We emphasise the influence of length scale $\ell$, the order of fractional continua $\alpha$, as well as the size of spatial discretization. The illustrative example comprises a one dimensional tension. The return-mapping algorithm is solved by analogy to the classical scheme discussed in \cite{Simo1997} Box 1.4.

\subsection{Geometry, boundary conditions and material parameters}

Total length of the body is $l=1~[m]$ (cf. Fig.~\ref{fig:example}). The displacements $U(X=0)=U(X=l)=\check{U}=0.003l$ causing tension are applied. The body force $b=615~[MNm^{-3}]$ is distributed through the body as shown in Fig.~\ref{fig:example}.

Material parameters are:
\begin{itemize}
	\item Young's modulus $E=205~[GPa]$,
	\item yield stress $\sigma_Y=1200~[MPa]$,
	\item[]and dependently on example
   \item the order of fractional continua $\alpha\in \{0.1,0.2,0.3,0.4,0.5,0.6,0.7,0.8,0.9,1.0\}$~[-],
	\item the length scale $\ell\in \{2\% l, 10\% l,20\% l\}~[m]$. 
\end{itemize}

\subsection{Numerical scheme}

Based on the discussion in Sec.\ref{ssec:BVP}, the analysed problem of one dimensional tension of elasto-plastic fractional continua with Dirichlet's boundary conditions is governed by the following flow chart

\begin{enumerate}
	\item Database at $X \in  \mathcal{B}:~\{ \varepsilon_k^p \}$.
	\item Compute elastic trial and test for plastic loading
	\begin{itemize}
			\item[] Given $\Delta\check{U}$ update:
			\begin{itemize}
				\item[] \textit{Displacements}
				\item[] $\frac{\partial}{\partial X}(\mathrm{Grad}\underset{X}{\tilde{\Delta {U}}})  + \frac{b}{E} =  0$;
				\item[] $U_{k+1}=U_k+\Delta U_k$;
				\item[]
				\item[] \textit{Strains}
				\item[] $\overset{\Diamond}{\varepsilon}_{k+1}=\overset{\Diamond}{\varepsilon}_k+\Delta\overset{\Diamond}{\varepsilon}_k$;
				\item[]
				\item[] \textit{Trial stress and yield function}
				\item[]$\sigma_{k+1}^{trial}=E(\overset{\Diamond}{\varepsilon}_{k+1}-\varepsilon_k^p)$;
				\item[]$f_{k+1}^{trial}=|\sigma_{k+1}^{trial}|-\sigma_Y$;
				\item[]
				\item[] {IF $f_{k+1}^{trial}\leq 0$ THEN}
				\item[] \textit{~~~~~~Elastic step}: set $(\cdot)_{k+1}=(\cdot)_{k+1}^{trial}$ $\&$ EXIT;
				\item[] {ELSE}
				\item[] \textit{~~~~~~Plastic step}: Proceed to step 3.
			\end{itemize}
	\end{itemize}
		\item Return mapping
			\begin{itemize}
			\item[] $\Delta \gamma=\frac{f_{k+1}^{trial}}{E}$;
			\item[] $\sigma_{k+1}=\sigma_{k+1}^{trial}-\Delta \gamma E \mathrm{sign}(\sigma_{k+1}^{trial})$;
			\item[] $\varepsilon_{k+1}^p=\varepsilon_k^p+\Delta \gamma \mathrm{sign}(\sigma_{k+1}^{trial})$.
			\end{itemize}
\end{enumerate}

The attention is paid to displacement increment calculation. Following definition of fractional derivative given by Eq.~(\ref{eq:RC}) for $\underset{X}{\tilde{\Delta{U}}}$ we need to calculate adequate left and right Caputo derivatives, so we have

\begin{equation}
\mathrm{Grad}\underset{X}{\tilde{\Delta{U}}} =\ell^{\alpha-1} ~_{~~a}^{RC}\underset{X}{D} ~^{\alpha}_b \Delta U=\ell^{\alpha-1} \frac{1}{2}\frac{\Gamma(2-\alpha)}{\Gamma(2)}\left(~_{a}^{C}D^{\alpha}_{X}\Delta U - ~_{X}^{C}D^{\alpha}_b\Delta  U \right).
\end{equation}

\begin{figure}[H]
\centering
\includegraphics[width=15cm]{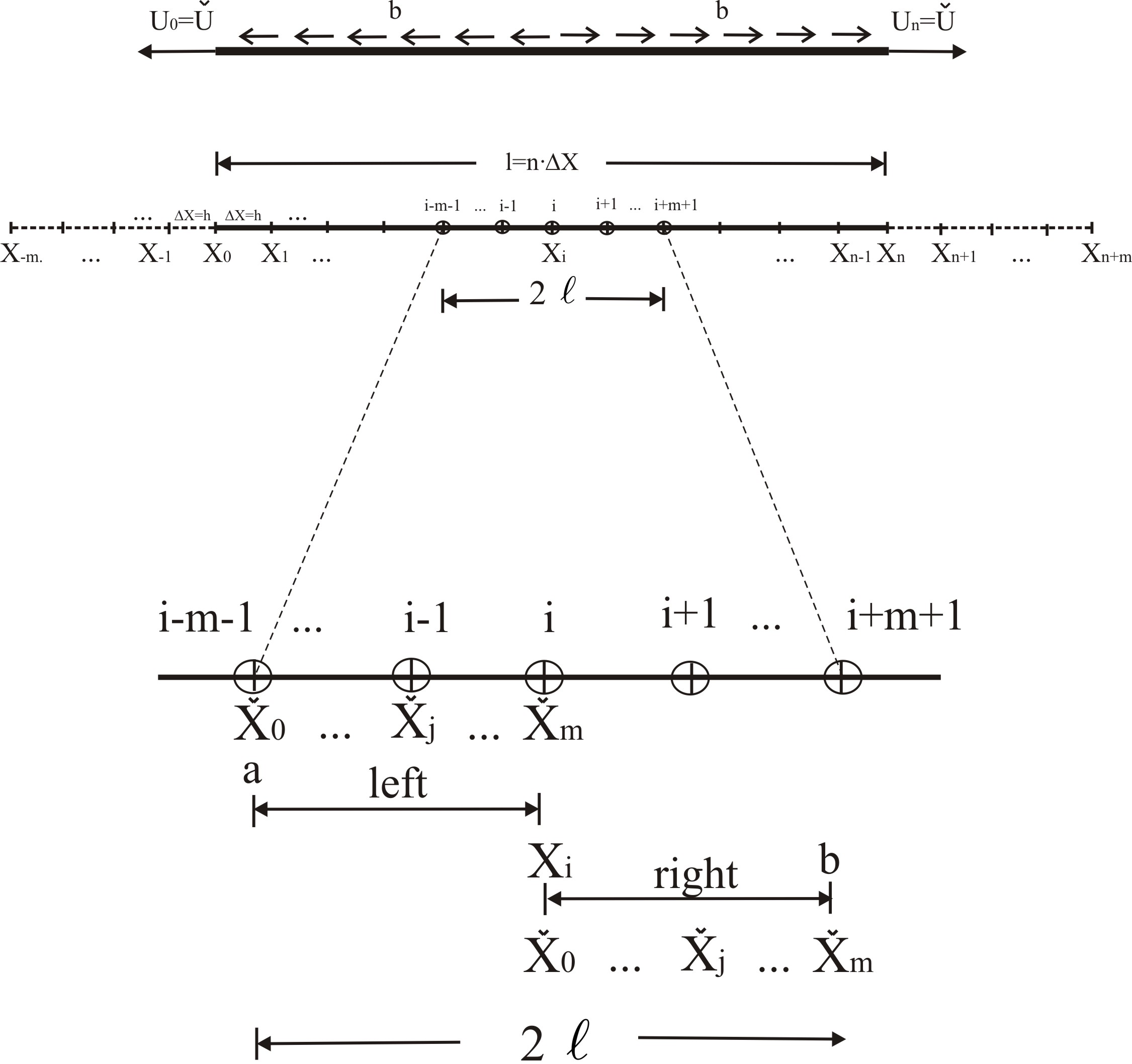}
\caption{Spatial discretization for one dimensional fractional continuum body}
\label{fig:example}
\end{figure}

For numerical calculations we propose the following approximation at the particular point of interest $X=X_i$ (Fig.~\ref{fig:example})
\begin{equation}\label{eq:genDM}
\frac{\partial}{\partial X}(\mathrm{Grad}\underset{X}{\tilde{\Delta{U}}})|_{X=X_i}\cong\frac{\mathrm{Grad}\underset{X}{\tilde{\Delta{U}}}|_i-\mathrm{Grad}\underset{X}{\tilde{\Delta{U}}}|_{i-1}}{\Delta X}.
\end{equation} 
The explicit formula for $\mathrm{Grad}\underset{X}{\Delta\tilde{{U}}}|_i$ utilising the modified trapezoidal rule can be defined as follows \cite{Odibat2006,Leszczynski2011}. 

For the left sided derivatives we use:
\begin{equation}
a=\check{X}_0<\check{X}_1<...<\check{X}_j<...<\check{X}_m=X, \quad h=\frac{\check{X}_m-\check{X}_0}{m}=\frac{X-a}{m}, \quad m \geq 2,
\end{equation}
\begin{eqnarray}\label{eq:leftDISC}
\nonumber
~_{a}^{C}D^{\alpha}_{t}\Delta U|_{X=\check{X}_m}\cong\frac{h^{n-\alpha}}{\Gamma(n-\alpha+2)}\left\{[(m-1)^{n-\alpha+1}-(m-n+\alpha-1)m^{n-\alpha}]\Delta U^{(n)}(\check{X}_0)\right.+\\
\Delta U^{(n)}(\check{X}_m)+\sum_{j=1}^{m-1}[(m-j+1)^{n-\alpha+1}-2(m-j)^{n-\alpha+1}+\left.(m-j-1)^{n-\alpha+1}] \Delta U^{(n)}(\check{X}_j)\right\},
\end{eqnarray}

where $\Delta U^{(n)}(\check{X}_j)$ denotes classical $n$-th derivative at $X=\check{X}_j$.

Similarly, for the right sided derivatives we use:
\begin{equation}
X=\check{X}_0<\check{X}_1<...<\check{X}_j<...<\check{X}_m=b, \quad h=\frac{\check{X}_m-\check{X}_0}{m}=\frac{b-X}{m}, \quad m \geq 2,
\end{equation}
\begin{eqnarray}\label{eq:rightDISC}
\nonumber
~_{t}^{C}D^{\alpha}_{b}\Delta U|_{X=\check{X}_0}\cong (-1)^n\frac{h^{n-\alpha}}{\Gamma(n-\alpha+2)}\left\{[(m-1)^{n-\alpha+1}-(m-n+\alpha-1)m^{n-\alpha}] \Delta U^{(n)}(\check{X}_m)\right.+\\
\Delta U^{(n)}(\check{X}_0)+\sum_{j=1}^{m-1}[(j+1)^{n-\alpha+1}-2j^{n-\alpha+1}+\left.(j-1)^{n-\alpha+1}]\Delta U^{(n)}(\check{X}_j)\right\}.
\end{eqnarray}

Notice that in limits $\alpha \shortrightarrow 0$ and $\alpha \shortrightarrow 1$ ($n=\alpha$), taking into account the approximations Eqs~(\ref{eq:leftDISC}) and (\ref{eq:rightDISC}), we obtain from Eq.~(\ref{eq:genDM}) the classical forward difference first-order, and central difference second-order, respectively.

Thus, the necessary information to be considered in approximations Eqs~(\ref{eq:leftDISC}) and (\ref{eq:rightDISC}) depends on chosen parameter $m$. As an example, to build the set of the linear equations, for calculation of displacements increment $\Delta U_{0},...,\Delta U_{n}$, for $m=2$, hence $n=1,~j=1,~h=\Delta X=\frac{\ell}{2}$ we have
\begin{equation}\nonumber
\frac{\partial}{\partial X}(\mathrm{Grad}\underset{X}{\tilde{\Delta{U}}})|_{X=X_i}\cong \frac{\mathsf{F}}{\Delta X^2}\left[\mathsf{B}\Delta U_{i-3}+(\mathsf{C}-2\mathsf{B})\Delta U_{i-2}+(\mathsf{B}-2\mathsf{C}+2)\Delta U_{i-1}+\right.
\end{equation}
\begin{equation}
\left. (\mathsf{C}+\mathsf{D}-4) \Delta U_i+(\mathsf{B}-2\mathsf{D}+2)\Delta U_{i+1}+(\mathsf{D}-2\mathsf{B})\Delta U_{i+2}+\mathsf{B} \Delta U_{i+3}\right],
\end{equation}
where
$$\mathsf{F}=\ell^{\alpha-1}\mathsf{E}\mathsf{A},$$
$$\mathsf{E}=\frac{1}{2}\frac{\Gamma(2-\alpha)}{\Gamma(2)},$$
$$\mathsf{A}=\frac{h^{n-\alpha}}{\Gamma(n-\alpha+2)}=\frac{h^{1-\alpha}}{\Gamma(3-\alpha)},$$
$$\mathsf{B}=(m-1)^{n-\alpha+1}-(m-n+\alpha-1)m^{n-\alpha}=1-\alpha 2^{1-\alpha},$$
$$\mathsf{C}=(m-j+1)^{n-\alpha+1}-2(m-j)^{n-\alpha+1}+(m-j-1)^{n-\alpha+1}=2^{2-\alpha}-2,$$
$$\mathsf{D}=(j+1)^{n-\alpha+1}-2j^{n-\alpha+1}+(j-1)^{n-\alpha+1}=2^{2-\alpha}-2.$$

By analogy for fractional strains we have:
\begin{itemize}
\item for $X_0$ we use forward difference for derivatives in  Eqs~(\ref{eq:leftDISC}) and (\ref{eq:rightDISC})

\begin{equation}\nonumber
\Delta \overset{\Diamond}{\varepsilon}=\mathrm{Grad}\underset{X}{\tilde{\Delta{U}}})|_{X=X_i}\cong \frac{\mathsf{F}}{\Delta X}\left[-\mathsf{B}\Delta U_{i-2}+(\mathsf{B}-\mathsf{C})\Delta U_{i-1}+\right.
\end{equation}
\begin{equation}
\left. (\mathsf{C}-2) \Delta U_{i}+(2-\mathsf{D})\Delta U_{i+1}+(\mathsf{D}-\mathsf{B})\Delta U_{i+2}+\mathsf{B}\Delta U_{i+3}\right],
\end{equation}

\item for $X_1\div X_{n-1}$ we use central difference for derivatives in  Eqs~(\ref{eq:leftDISC}) and (\ref{eq:rightDISC})

\begin{equation}\nonumber
\Delta \overset{\Diamond}{\varepsilon}=\mathrm{Grad}\underset{X}{\tilde{\Delta{U}}})|_{X=X_i}\cong \frac{\mathsf{F}}{2\Delta X}\left[-\mathsf{B}\Delta U_{i-3}-\mathsf{C} \Delta U_{i-2}+(\mathsf{B}-2) \Delta U_{i-1}+\right.
\end{equation}
\begin{equation}
\left. (\mathsf{C}-\mathsf{D})\Delta U_i+(2-\mathsf{B})\Delta U_{i+1}+\mathsf{D}\Delta U_{i+2}+\mathsf{B}\Delta U_{i+3}\right],
\end{equation}

\item for $X_{n}$ we use backward difference for derivatives in  Eqs~(\ref{eq:leftDISC}) and (\ref{eq:rightDISC})

\begin{equation}\nonumber
\Delta \overset{\Diamond}{\varepsilon}=\mathrm{Grad}\underset{X}{\tilde{\Delta {U}}})|_{X=X_i}\cong \frac{\mathsf{F}}{\Delta X}\left[-\mathsf{B}\Delta U_{i-3}+(\mathsf{B}-\mathsf{C})\Delta U_{i-2}+\right.
\end{equation}
\begin{equation}
\left. (\mathsf{C}-2)\Delta U_{i-1}+(2-\mathsf{D})\Delta U_{i}+(\mathsf{D}-\mathsf{B})\Delta U_{i+1}+\mathsf{B} \Delta U_{i+2}\right].
\end{equation}

\end{itemize}

Thus spatial discretization of the analysed problem is as shown in Fig.~\ref{fig:example}. We have the physical nodes $X_0,...,X_n$ and fictitious nodes on the left $X_{-m},...,X_{-1}$, and the right $X_{n+1},...,X_{n+m}$. The fictitious nodes are due to the definition of fractional derivative operator on an interval. We assume, by analogy as in \cite{Ciesielski2005}, that for all fictitious nodes on the left the displacements are $U_{-m},...,U_{-1}=U(X=0)$ while for all fictitious nodes on the right the displacements are  $U_{n+1},...,U_{n+m}=U(X=l)$. In Fig.~\ref{fig:example} it is also shown, that in general, for point of interest $X_i$, one needs to govern the information from $2m+3$ points (recall that $\ell$ equals the half of the integration interval $(a,b)$ and the point of interest lays in the middle of this interval).

\vspace{.5mm}

\subsection{Discussion on plastic strains}

\begin{figure}[H]
\centering
\includegraphics[width=15cm]{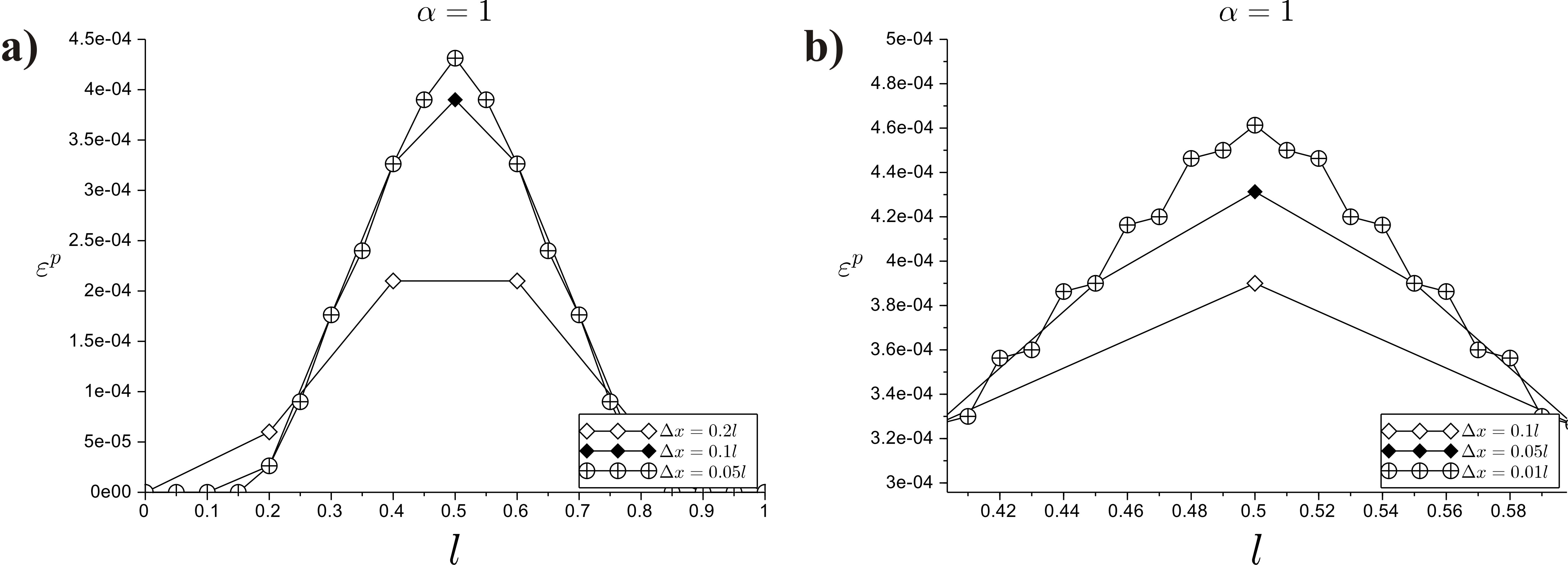}
\caption{Classical solution of the plastic strains distribution through the length of the body for different spatial discretization: a) whole body; b) magnification of the pick plastic strains zone.}
\label{fig:R1}
\end{figure}

In Fig.~\ref{fig:R1} the classical solution is presented. As mentioned this special case is obtained for $\alpha=1$, and length scale $\ell$ does not influence the results ($\ell^{1-\alpha}=\ell^{1-1}=\ell^0=1$). We observe that classical solution (as a local one) is sensitive on a spatial discretization (the size of $\Delta X$). We see, that for $\Delta X =0.2l$ the plastic zone covers whole body, and pick plastic strains are twice smaller than for finer discretizations.

In Fig.~\ref{fig:R1} we observe how the applied length scale and order of fractional continua influence the distribution of plastic strains. The results are obtained for spatial discretization $\Delta X = \frac{\ell}{2}$ ($m=2$) - let us point out that one can not apply coarser one due to discretization scheme discussed in previous section. One should notice that the presented non-local fractional formulation allows in a very flexible way control the dimension of plastic strains zone as well as their magnitude. Is is important that when $\ell\shortrightarrow 0$, most of the results converge to classical local formulation. On the other side please notice that there can exist the orders of fractional continua for which the solution is beyond engineering intuition (cf. results for $\alpha=0.2$).

\begin{figure}[H]
\centering
\includegraphics[width=15cm]{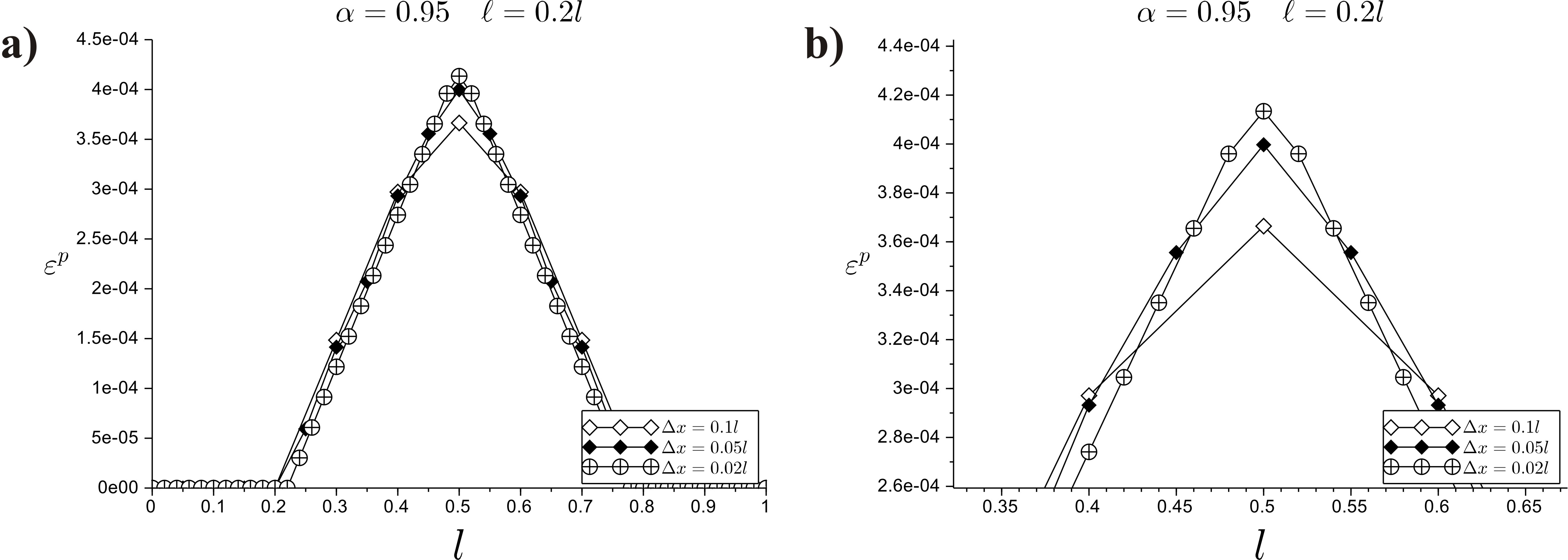}
\caption{Plastic strains distribution through the length of the body for $\alpha=0.95$ and $\ell=0.2l$ for different spatial discretization: a) whole body; b) magnification of the pick plastic strains zone.}
\label{fig:R3-b}
\end{figure}

\begin{figure}[H]
\centering
\includegraphics[width=15cm]{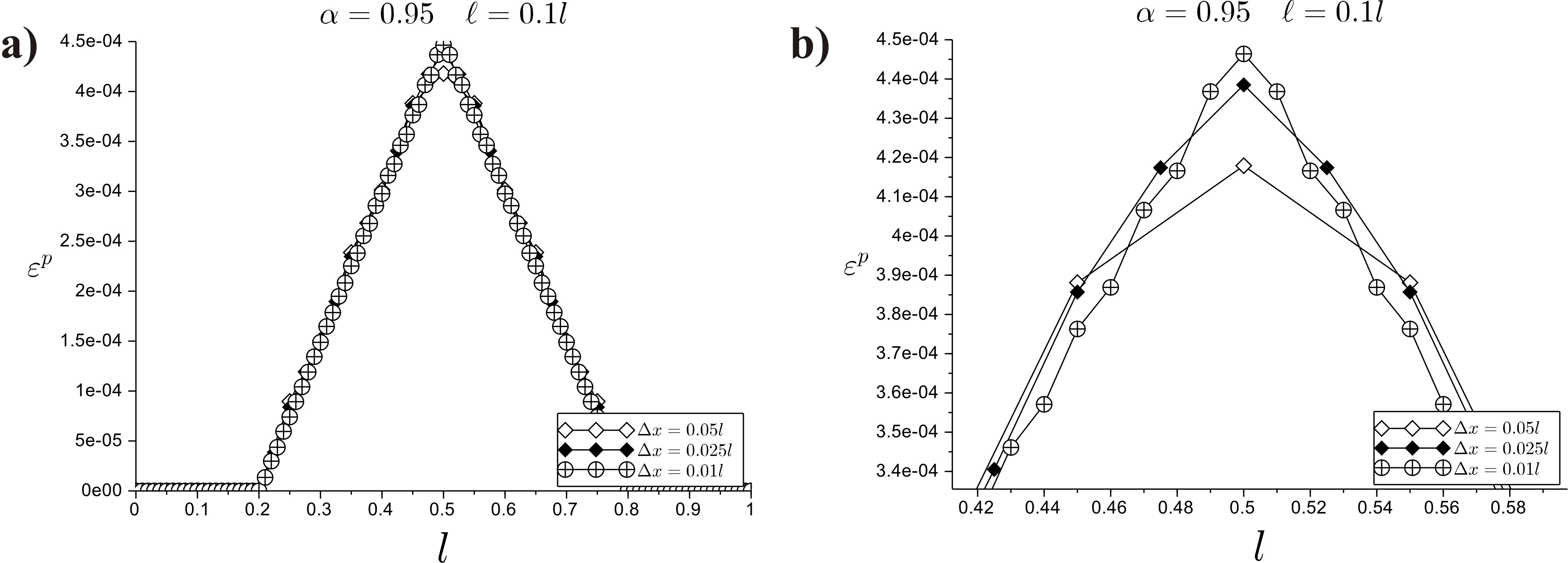}
\caption{Plastic strains distribution through the length of the body for $\alpha=0.95$ and $\ell=0.1l$ for different spatial discretization: a) whole body; b) magnification of the pick plastic strains zone.}
\label{fig:R3-a}
\end{figure}

\begin{figure}[H]
\centering
\includegraphics[width=15cm]{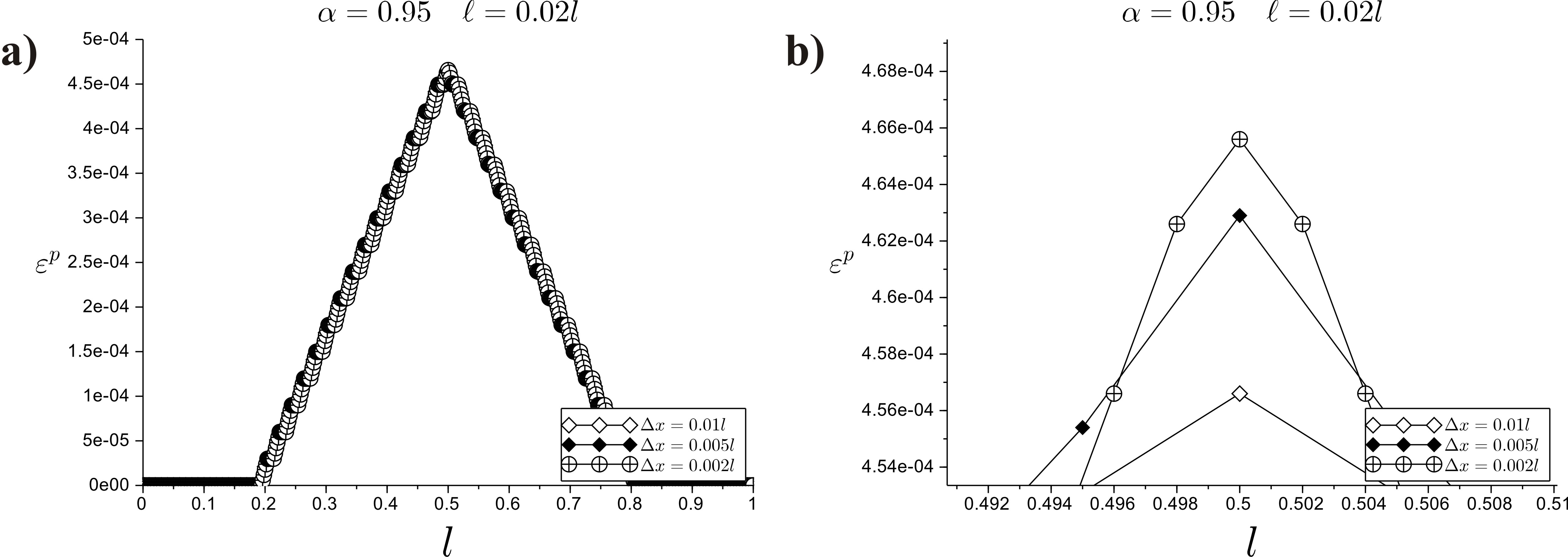}
\caption{Plastic strains distribution through the length of the body for $\alpha=0.95$ and $\ell=0.02l$ for different spatial discretization: a) whole body; b) magnification of the pick plastic strains zone.}
\label{fig:R3-c}
\end{figure}

\begin{figure}[H]
\centering
\includegraphics[width=15cm]{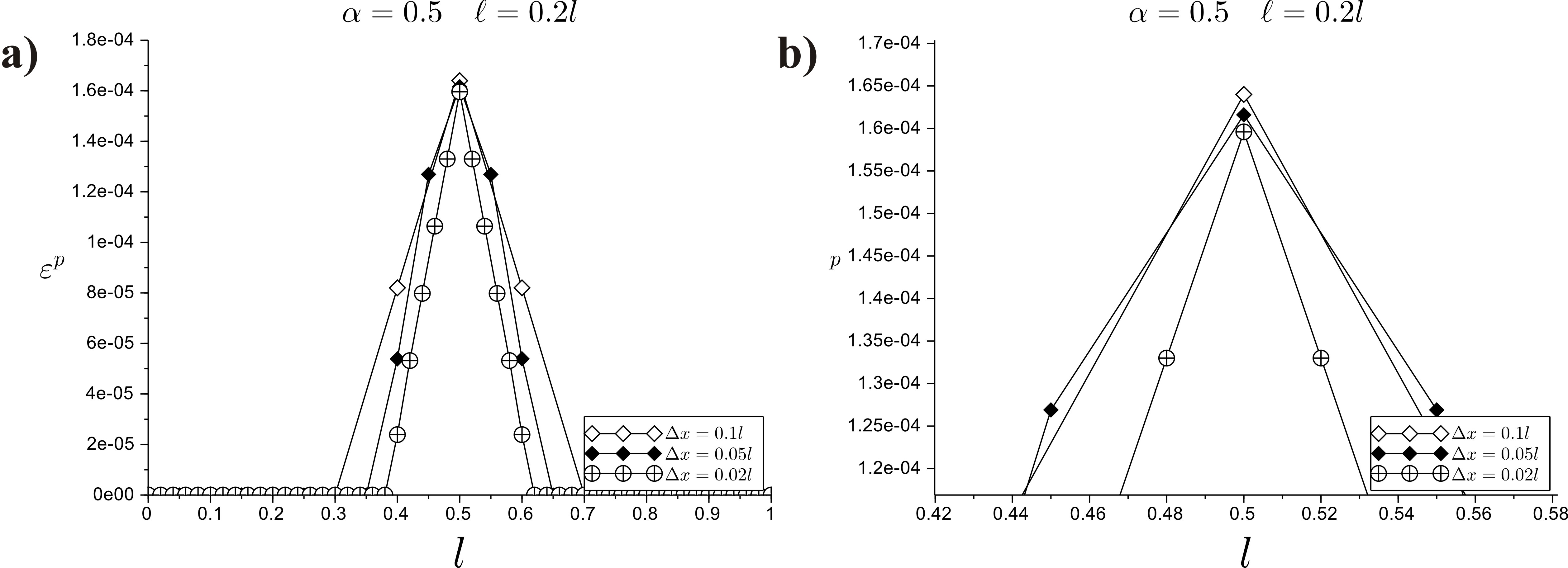}
\caption{Plastic strains distribution through the length of the body for $\alpha=0.5$ and $\ell=0.2l$ for different spatial discretization: a) whole body; b) magnification of the pick plastic strains zone.}
\label{fig:R4-b}
\end{figure}

\begin{figure}[H]
\centering
\includegraphics[width=15cm]{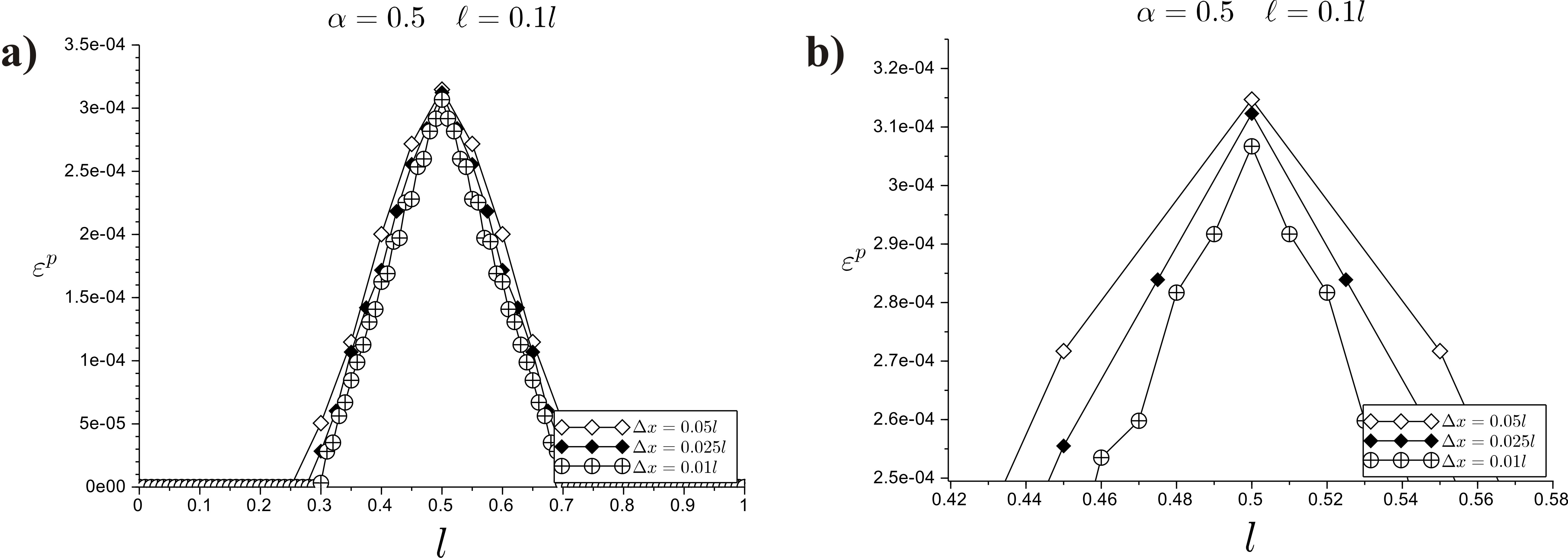}
\caption{Plastic strains distribution through the length of the body for $\alpha=0.5$ and $\ell=0.1l$ for different spatial discretization: a) whole body; b) magnification of the pick plastic strains zone.}
\label{fig:R4-a}
\end{figure}

\begin{figure}[H]
\centering
\includegraphics[width=15cm]{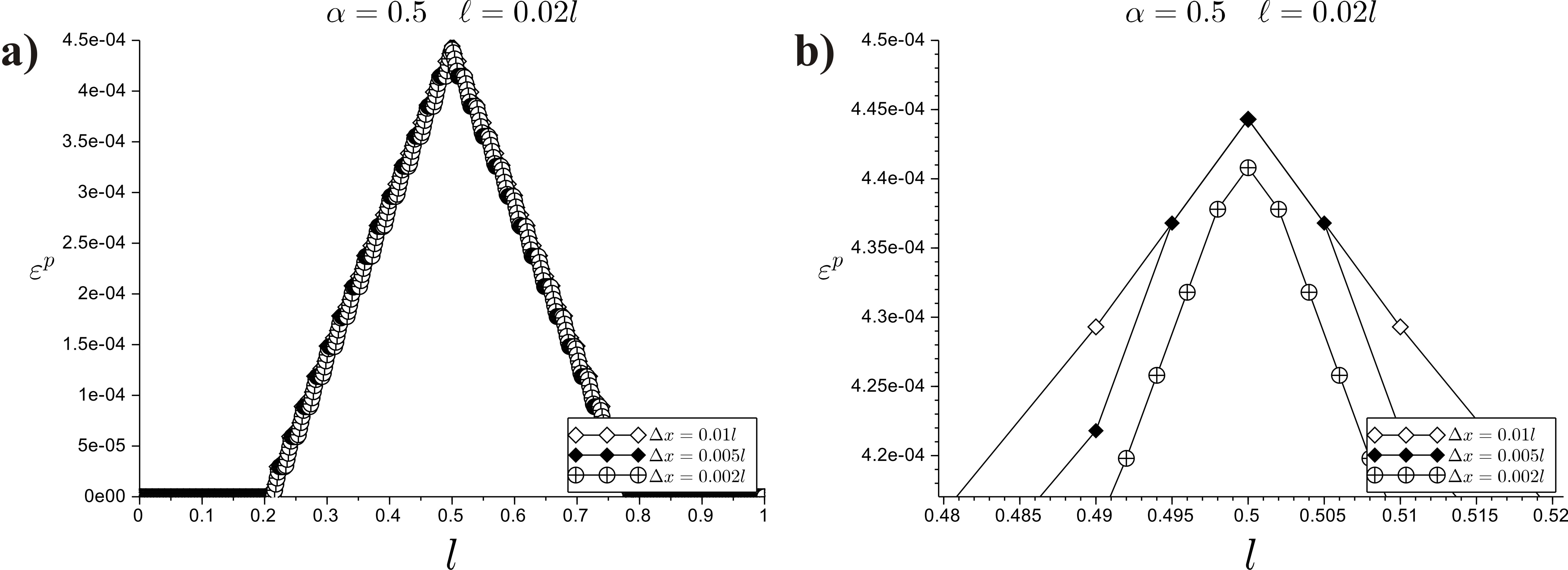}
\caption{Plastic strains distribution through the length of the body for $\alpha=0.5$ and $\ell=0.02l$ for different spatial discretization: a) whole body; b) magnification of the pick plastic strains zone.}
\label{fig:R4-c}
\end{figure}

\begin{figure}[H]
\centering
\includegraphics[width=7cm]{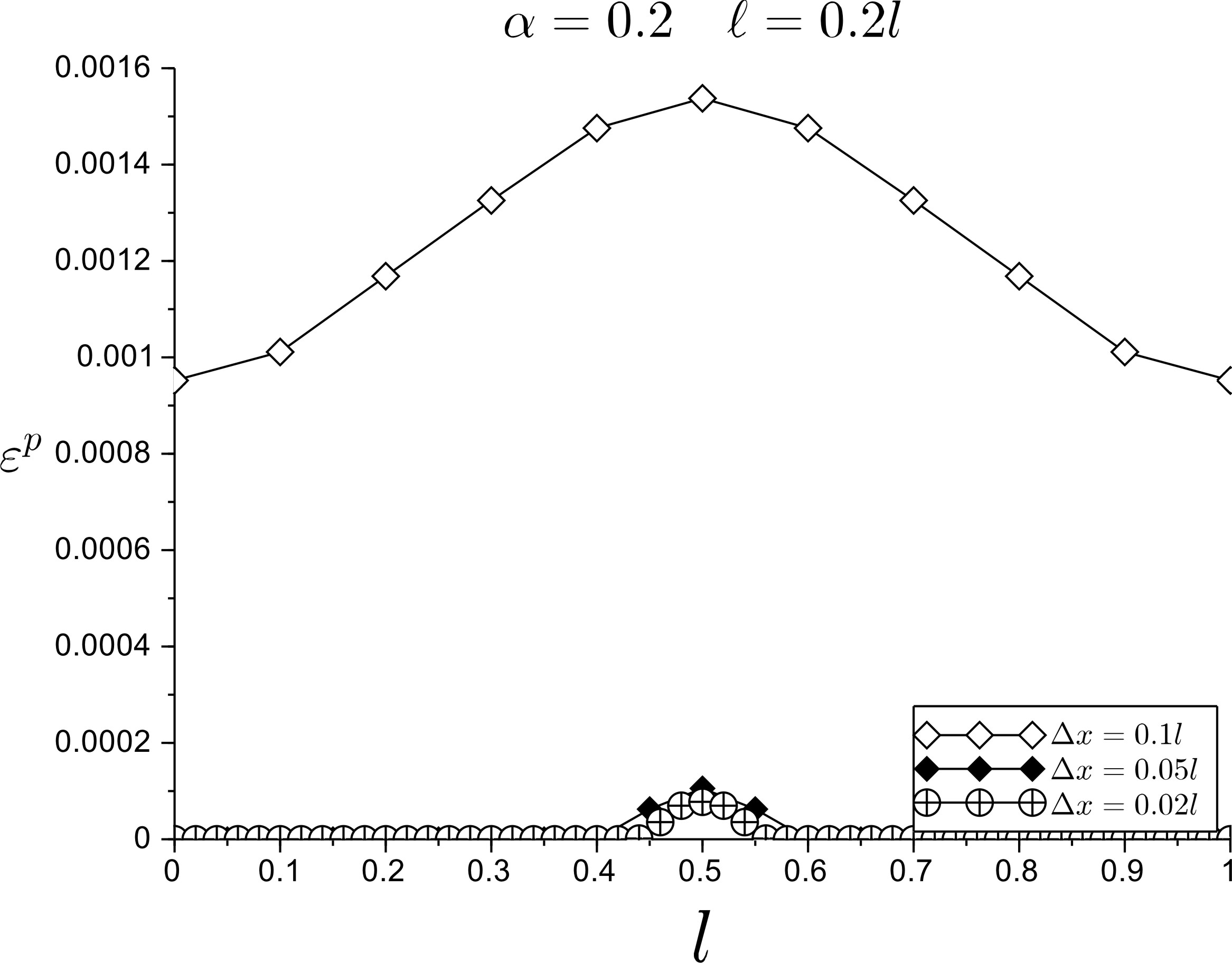}
\caption{Plastic strains distribution through the length of the body for $\alpha=0.2$ and $\ell=0.2l$ for different spatial discretization - whole body.}
\label{fig:R5-b}
\end{figure}

\begin{figure}[H]
\centering
\includegraphics[width=7cm]{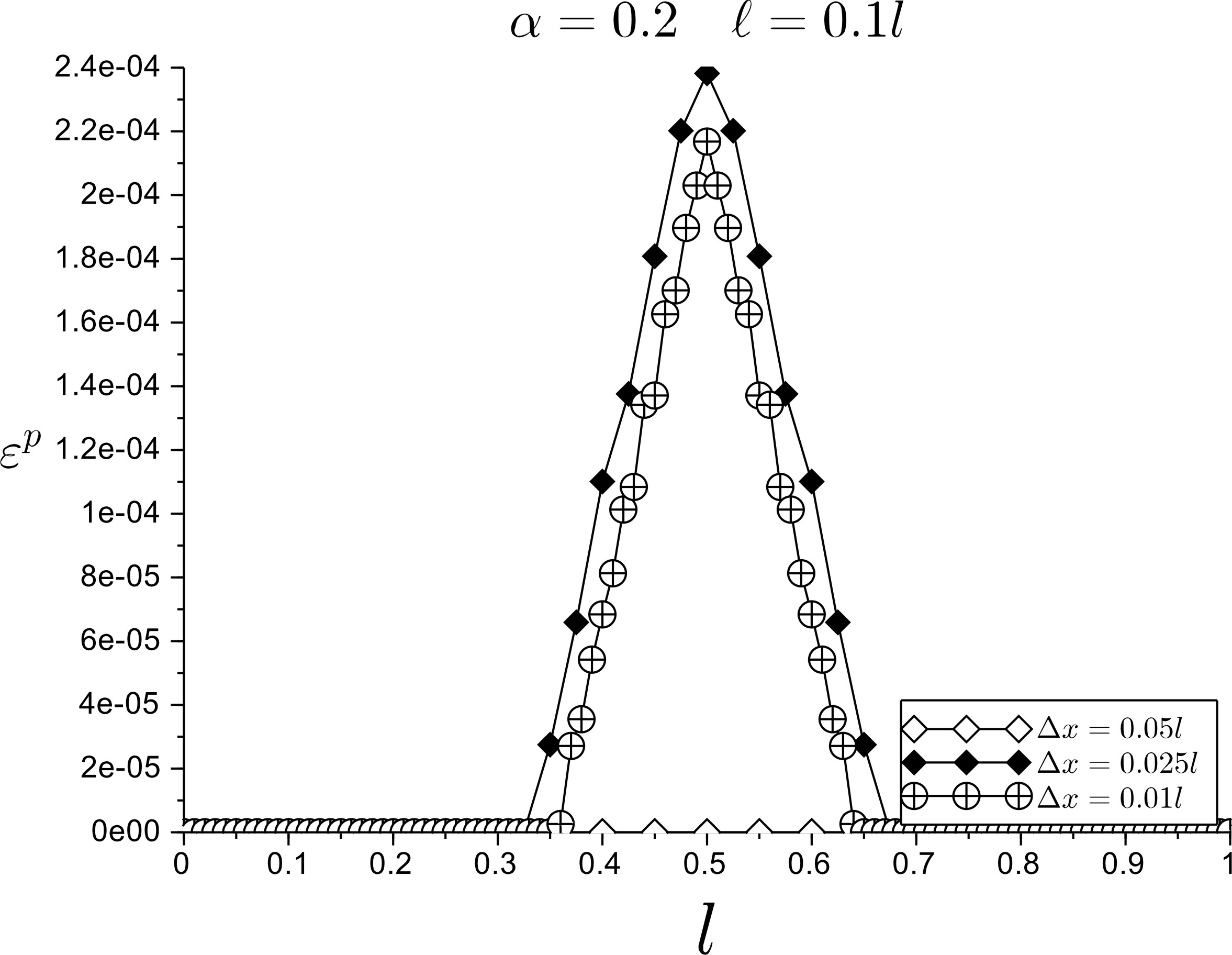}
\caption{Plastic strains distribution through the length of the body for $\alpha=0.2$ and $\ell=0.1l$ for different spatial discretization - whole body.}
\label{fig:R5-a}
\end{figure}

\begin{figure}[H]
\centering
\includegraphics[width=15cm]{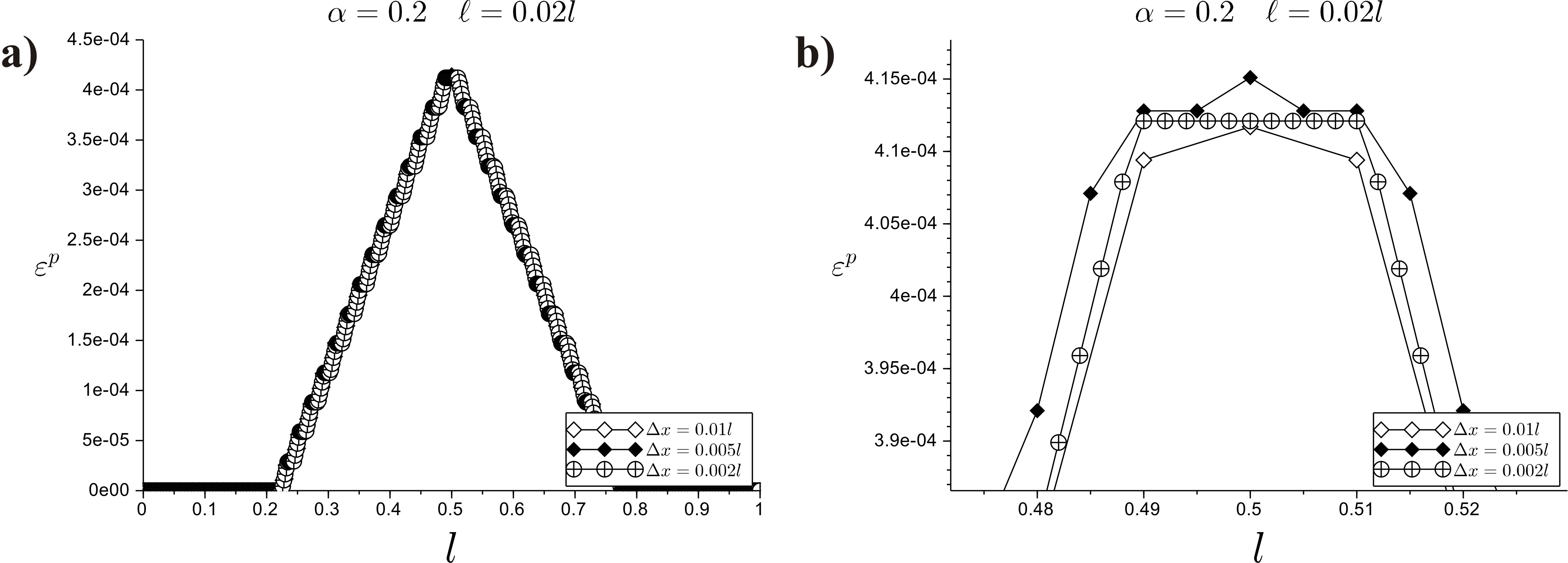}
\caption{Plastic strains distribution through the length of the body for $\alpha=0.2$ and $\ell=0.02l$ for different spatial discretization: a) whole body; b) magnification of the pick plastic strains zone.}
\label{fig:R5-c}
\end{figure}

In Figs~\ref{fig:R3-b}$\div$\ref{fig:R5-c} the influence of spatial discretization on the non-local results are presented. Nine cases are considered:
\begin{itemize}
	\item $\alpha=0.95$, $\ell=0.2l$, and $m\in \{2,4,10\}$ (Fig.~\ref{fig:R3-b}),
	\item $\alpha=0.95$, $\ell=0.1l$, and $m\in \{2,4,10\}$ (Fig.~\ref{fig:R3-a}),
	\item $\alpha=0.95$, $\ell=0.02l$, and $m\in \{2,4,10\}$ (Fig.~\ref{fig:R3-c}),
	\item $\alpha=0.5$, $\ell=0.2l$, and $m\in \{2,4,10\}$ (Fig.~\ref{fig:R4-b}),
	\item $\alpha=0.5$, $\ell=0.1l$, and $m\in \{2,4,10\}$ (Fig.~\ref{fig:R4-a}),
	\item $\alpha=0.5$, $\ell=0.02l$, and $m\in \{2,4,10\}$ (Fig.~\ref{fig:R4-c}),
	\item $\alpha=0.2$, $\ell=0.2l$, and $m\in \{2,4,10\}$ (Fig.~\ref{fig:R5-b}),
	\item $\alpha=0.2$, $\ell=0.1l$, and $m\in \{2,4,10\}$ (Fig.~\ref{fig:R5-a}),
	\item $\alpha=0.2$, $\ell=0.02l$, and $m\in \{2,4,10\}$ (Fig.~\ref{fig:R5-c}).
\end{itemize}
We observe that introduction of fractional non-locality (as in classical non-local models) regularises the results. We observe that the plastic strains distribution is not so sensitive on spatial discretization - in contrast to classical solution cf. Fig.~\ref{fig:R1}. Nevertheless, once more, we see strange solution for $\alpha=0.2$ - in this sense there exist a limit value for the order of fractional continua applicable for a specific phenomena/process being modelled. 

\begin{figure}[H]
\centering
\includegraphics[width=15cm]{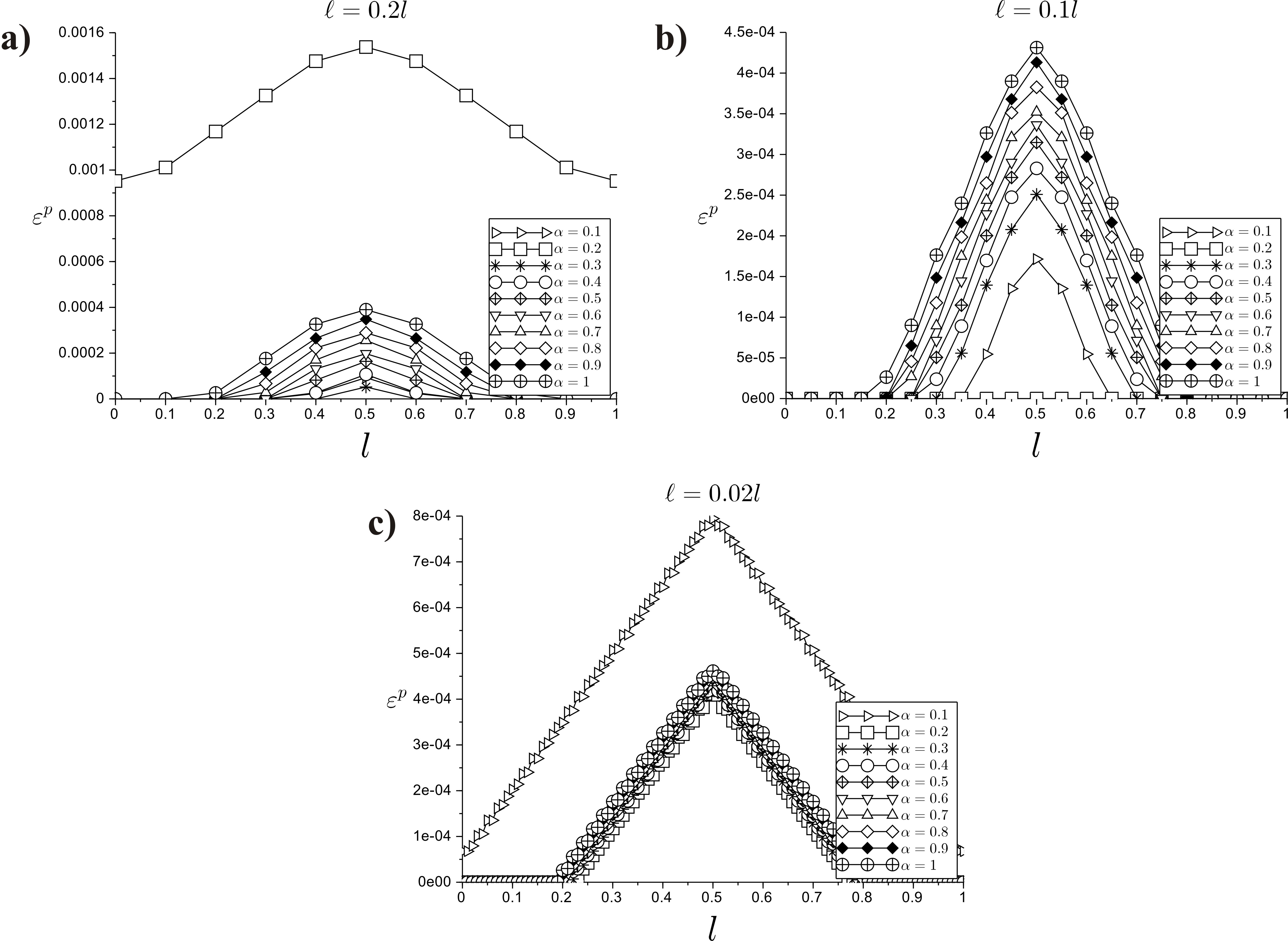}
\caption{Plastic strains distribution through the length of the body versus different length scales and orders of fractional continua.}
\label{fig:R2}
\end{figure}

\section{Conclusions}

Non-local fractional model of rate independent plasticity introduces a flexible tool for modelling elasto-plasic bodies with only two additional material parameters in comparison to the classical model: length scale, and order of fractional continua. Based on one the dimensional tension example we have shown that using the proposed formulation we can control the distribution and the magnitude of plastic strains. The insensitivity to the spatial discretization is presented also. On the other side, it should be emphasised that using this new formulation  one can obtain the solution which is beyond engineering intuition.

Finally, recall the genius sentence by Leibniz \cite{Leibniz62} about the fractional calculus, perspectives: \textit{"It will lead to a paradox, from which one day useful consequences will be drawn"}.

\bibliographystyle{apalike}
\bibliography{Biblio-Sumelka}

\end{document}